\documentclass{cernrep} 
\usepackage{texnames} \usepackage{epstopdf} \usepackage{tabularx} \usepackage[T1]{fontenc}
\usepackage{booktabs}
\usepackage[bookmarks, colorlinks=true, linktoc=page, pdftex, linkcolor=black, citecolor=black, urlcolor=blue]{hyperref}
\newcommand{\RM}[1]{\mathrm{#1}}
 
\newcommand{\avg}[1]{\left< #1 \right>} 
\renewcommand{\d}[2]{\frac{d #1}{d #2}} 
 
\usepackage{fancyhdr}
\fancyhfoffset{4 mm}
\fancypagestyle{ARTTITLE}{%
\fancyhf{} 
\lhead{\small{Proceedings of the CERN--Accelerator--School course: \it{Introduction to Accelerator Physics}}} 
\lfoot{Available online at \url{https://cas.web.cern.ch/previous-schools}}
\rfoot{\thepage\hspace*{3mm}}
 
}

\begin{document}
\title{Cyclotrons and Fixed Field Alternating Gradient Accelerators}
 
\author {M.\,Seidel}

\institute{Paul Scherrer Institut and \'Ecole Polytechnique F\'ed\'erale Lausanne, Switzerland}

\begin{abstract}
Due to its simplicity the classical cyclotron has been used very early for applications in science, medicine and industry. Higher energies and intensities were achieved through the concepts of the sector focused isochronous cyclotron and the synchro-cyclotron. Besides those the fixed field alternating gradient accelerator (FFA) represents the most general concept among these types of fixed field accelerators, and the latter one is actively studied and developed for future applications. 
\end{abstract}

\keywords{Cyclotron; FFA; fixed-field; isochronous.}

\maketitle 
\thispagestyle{ARTTITLE}

\section{Introduction} \label{sec:into} 
 
Cyclotrons have a long history in accelerator physics and are used for a wide range of medical, industrial, and research applications \cite{onishenko, seidel_cal}. The first cyclotrons were designed and built by Lawrence and Livingston \cite{lawrence, livingston} back in 1931. The cyclotron represents a resonant-accelerator concept. Due to the~repeated acceleration process it is possible to achieve relatively high kinetic energies while the~required voltages stay in a moderate range. Historically that was a major advantage over single pass high voltage accelerators. Furthermore, several properties such as CW operation make the concept well suited for the~acceleration of hadron beams with high average intensity. Relativistic effects limit the energies in reach for the classical cyclotron concept. These limitations were overcome to some degree by the~sector focused isochronous cyclotron, that allows to accelerate protons to a range of $\approx 1\,$GeV. Another approach for reaching such energies is the synchro-cyclotron, that involves cycling of the RF frequency at the expense of a lower average beam intensity. Scaling and non-scaling FFA utilize strong focusing to reduce the radial orbit variation. In a sense FFA concepts represent a generalisation of the cyclotron. As a major difference to synchrotrons the magnetic fields of FFA's and cyclotrons are not cycled during acceleration, and thus higher average intensities can be achieved.

\section{Cyclotron Concepts} \label{sec:classical}

\begin{figure} 
\centering\includegraphics[width=0.65\textwidth]{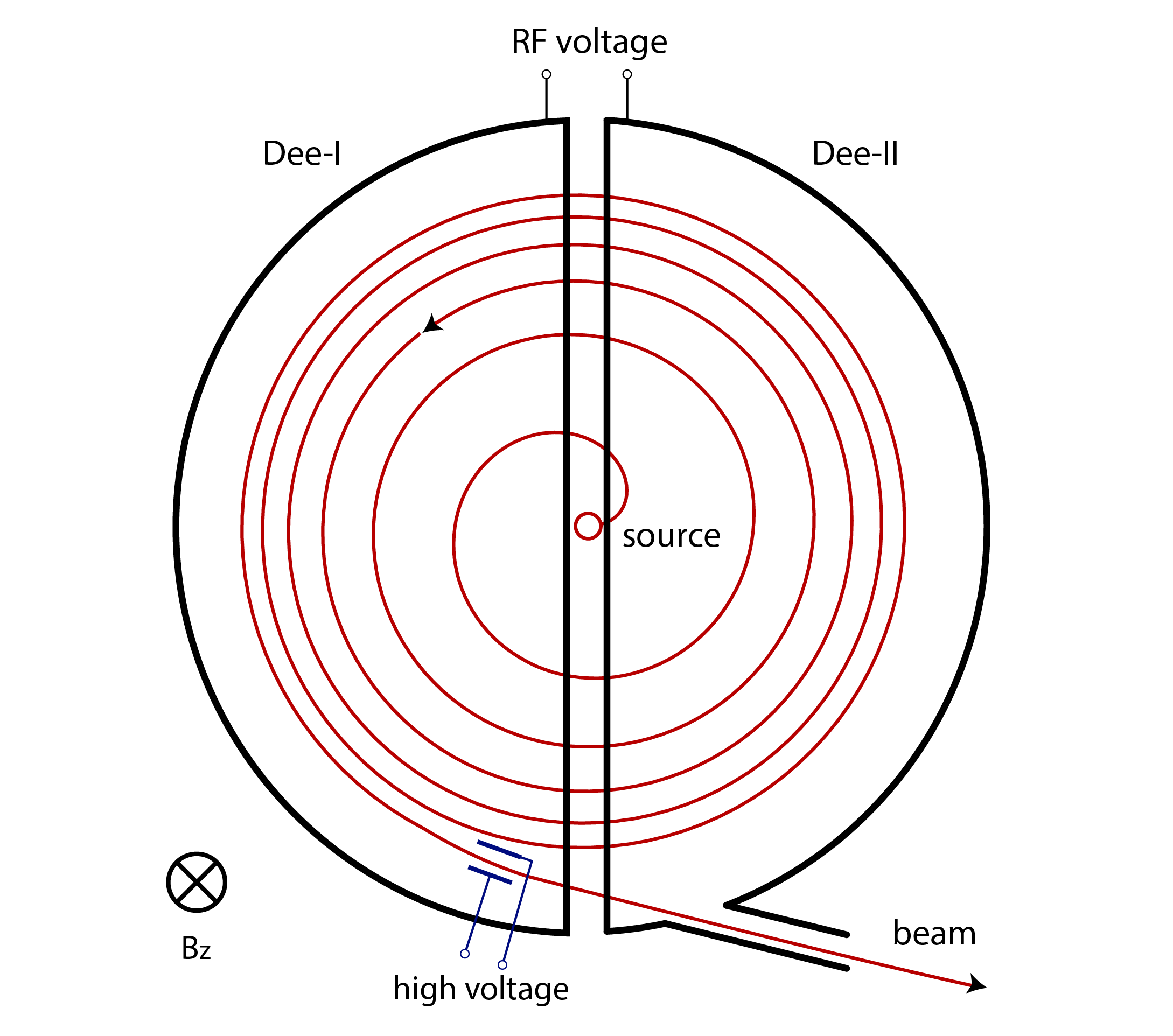}
\caption{Conceptual sketch of a classical cyclotron in plan view. In the non-relativistic approximation, the turn separation scales as $1/R$.}
\label{fig:spiral}
\end{figure}

Although the classical cyclotron has major limitations and is practically outdated today, some fundamental relations are best explained within this original concept. In the classical cyclotron an alternating high voltage at radio frequency (RF) is applied to two D-shaped hollow electrodes, the \emph{Dees}, for the purpose of acceleration. Ions from a central ion source are repeatedly accelerated from one dee to the other. The~ions are kept on a piecewise circular path by the application of a uniform, vertically oriented magnetic field. On the last turn, the ions are extracted by applying an electrostatic field using an electrode. The~concept is illustrated in Fig.~\ref{fig:spiral}.

The equation of motion for a particle moving in a vertically oriented magnetic field with a momentum vector in the horizontal plane is given by:

\begin{equation}
m \ddot{\vec{r}} = m\,\omega^2 \vec{r} = q\,\dot{\vec{r}}\times \vec{B}_z.
\label{cfug}
\end{equation}

The magnetic force and centrifugal force are set equal. The solution for the particles trajectory is a circle with a bending radius $\rho$ that is constant over time:
\begin{align*}
\vec{r}(t) = \rho \left( \begin{array}{c} \cos\omega t\\ \sin\omega t\\ 0 \end{array} \right).
\end{align*}

For kinetic energies low compared to the rest energy the particles circulate at the \emph{cyclotron frequency}, which depends on the magnetic field $B_z$, the charge $q$ and the rest mass $m_0$ of the particles:

\begin{eqnarray} \omega_c & = & \frac{qB_z}{m_0}
\nonumber \\[5pt]
f_c = \frac{\omega_c}{2\pi} & \approx & 15.2~ \RM{MHz}\cdot B_z \RM{[T]}
\RM{~(\,for\,\,protons\,)}.
\label{omega_c}
\end{eqnarray}

As soon as relativistic effects become important, the revolution frequency decreases with the relativistic factor $\gamma$ as: 
\begin{equation}
\omega_\RM{rev} = \omega_c/\gamma \,. 
\label{omega_rev}
\end{equation}

The bending radius $\rho$ is given by the well known bending strength for charged particles:

\begin{eqnarray} 
B\rho\,[\RM{T\cdot m}] & = & 3.336 \cdot \beta \cdot E_\RM{tot}\,[\RM{GeV}] \nonumber \\[5pt]
&\approx & 4.426 \cdot \sqrt{E_k/m_oc^2}\, .
\end{eqnarray}

The relation is expressed in practical units and the second line is an approximation for particles with kinetic energies much lower than their rest energy. $E_\RM{tot}$ is the sum of kinetic energy $E_k$ and rest energy $m_0 c^2$. The frequency of the accelerating voltage must be equal to the revolution frequency or an~integer multiple of it, i.e., $\omega_\RM{rf} = h \omega_\RM{rev}$. The harmonic number $h$ corresponds to the number of bunches that can be accelerated in one turn. With increasing velocity, particles travel at larger radii, so that $R \propto \beta$, and the revolution time remains constant and in phase with the RF voltage. We denote the average orbit radius with $R$, which may differ from the local bending radius $\rho$ for certain field configurations. In the~literature on cyclotrons the variable $R_\infty$ has been introduced to parametrize the~dependence of the~orbit radius on the particles velocity:

\begin{align}
R = R_\infty \beta, ~R_\infty = \frac{c}{\omega_c}\, .
\label{r_infty}
\end{align}

$R_\infty$ is a theoretical value, the trajectory radius for a particle at infinite energy and speed of light. For low energies the condition of \emph{isochronicity} is fulfilled in a homogeneous magnetic field. For relativistic particles the magnitude of the B-field has to be raised in proportion to $\gamma$ at increasing radius in order to keep the revolution time constant throughout the acceleration process. Such field shape is introduced for isochronous cyclotrons as described later in Section \ref{sec:avf}. To summarize this important result - for constant revolution frequency in a cyclotron the following scaling of orbit radius and bending field is required:

\begin{equation} R \propto \beta,~~B_z \propto \gamma \, .
\label{isochron}
\end{equation}

As we will see this field scaling contradicts the requirements of transverse focusing. The radial variation of the bending field in a cyclotron generates focusing forces. At a radius $R$, the slope of the~bending field is described by the field index\footnote{In the literature also the variable $n=-k$ is used for the field index.} $k$, where

\begin{equation} 
k = \frac{R}{B_z} \d{B_z}{R}.
\label{index} 
\end{equation} 

Using the proportionalities Eq. (\ref{isochron}), the scaling of the field index under isochronous conditions can be evaluated as follows:

\begin{eqnarray}
\frac{R}{B} \frac{\RM{d}B}{\RM{d}R} & = &
\frac{\beta}{\gamma} \frac{\RM{d}\gamma}{\RM{d}\beta}
\nonumber \\[5pt]
& = & \gamma^2 -1.
\label{index1}
\end{eqnarray}

Note that Eq. (\ref{index1}) is positive, thus isochronicity requires an increasing magnetic field towards larger radii. The radial equation of motion of a single particle can be written as

\begin{equation} 
m\ddot{r} = m r \dot{\varphi}^2-q r\dot{\varphi} B_z .
\end{equation}

We now consider small deviations around the central orbit $R$, namely $r=R+x,x \ll R$:

\begin{eqnarray} 
\ddot{x} + \frac{q}{m} v B_z(R+x) - \frac{v^2}{R+x} & = & 0, \nonumber \\[6pt] 
\ddot{x} + \frac{q}{m} v \left( B_z(R) + \frac{\RM{d}B_z}{\RM{d}R} x\right) 
-\frac{v^2}{R} \left(1-\frac{x}{R} \right) & = & 0, \nonumber \\[6pt]
\ddot{x} + \omega_\RM{rev}^2 (1+k) x & = & 0. 
\label{radmotion}
\end{eqnarray}
In this derivation, we have used the relations $\omega_\RM{rev} = qB_z/m \approx v/R$ and $r\dot{\varphi} \approx v$. Thus, in the linear approximation, the horizontal `betatron motion' is a harmonic oscillation around the central beam orbit, $x(t) = x_\RM{max} \cos(\nu_r \omega_\RM{rev} t)$. The parameter $\nu_r$ is called the betatron tune. From Eq. (\ref{radmotion}) we see that the~radial betatron tune in a classical cyclotron is given by

\begin{align}
\nu_r & = \sqrt{1+k}\, .
\label{nu_r}
\end{align}

For an isochronous cyclotron, using Eq. (\ref{index1}), we obtain in addition:

\begin{align}
\nu_r & \approx \gamma.
\label{nu_r_gamma}
\end{align}

A similar calculation can be done for the vertical plane, using Maxwell's equation $\RM{rot} \, \vec{B}=0$. This yields for the vertical betatron frequency:

\begin{equation}
\nu_z = \sqrt{-k}.
\label{nu_z}
\end{equation}

As is obvious from Eq. (\ref{nu_z}), vertical focusing can be obtained only if the bending field decreases towards larger radii, i.e. $k<0$. However, a negative slope of the field would be inconsistent with the~condition of isochronicity as stated above, which requires the field to increase in proportion to $\gamma$. Classical cyclotrons use a zero or slightly negative field index and their energy reach is limited by the~resulting phase slip. In Section \ref{sec:synchro} we illustrate acceleration with imperfect synchronism in a numerical example. As we will see in Section \ref{sec:avf}, vertical focusing can in fact be achieved by an azimuthal variation of the~bending field.

Also for cyclotrons a beam optical beta-function can be defined in the sense of the Courant--Snyder theory \cite{courant}. In the radial plane, the average beta function can be estimated via

\begin{equation}
\beta_r \approx \frac{R}{\nu_r}.
\label{beta}
\end{equation}

In practice $\nu_r \approx \gamma$ is a small number with little room for adjustments. The beam size in cyclotrons, scaling with $\sqrt{\beta_r}$, is therefore coupled to the orbit radius and the size of the cyclotron. This is a major difference to the strong focusing accelerator optics as used in synchrotron and FFA, for which the tune scales with the number of lattice cells and thus with radius. For those machines the optics is a function of the lattice cell and is practically decoupled from the size of the accelerator. A radial dispersion function can be defined as well:

\begin{eqnarray}
D_r & \equiv & \frac{\Delta R}{\Delta p/p} \label{dispersion}\, , \nonumber \\[5pt]
D_r & \approx & \frac{R}{\nu_r^2} \approx \frac{R}{\gamma^2}. 
\label{disp_apx}
\end{eqnarray}

Similar relations are obtained for synchrotrons using the so called smooth approximation. The~derivation of Eq. (\ref{disp_apx}) is done in a similar way as the calculation of the radial step width in Eq.~(\ref{step_1}) in the next section. The two relations for $\beta_r$ and $D_r$ can be used to establish rough matching conditions for beam injection into cyclotrons. 
 
The so-called $K$-value is a commonly used parameter for the characterization of the magnetic energy reach of a cyclotron design. It equals the maximum attainable energy for single charged particles with a mass of $1/12$ of a Carbon nucleus (roughly a proton) in the non-relativistic approximation. The~$K$-value is proportional to the maximum squared bending strength, i.e., $K \propto (B\rho)^2$, and can be used to rescale the achievable kinetic energy per nucleon for varying charge-to-mass ratio:

\begin{equation} \frac{E_\RM{k}}{A} = K \left( \frac{Q}{A} \right)^2.
\end{equation}

The original cyclotron concept exhibits essential properties that allow the use of cyclotrons for high-intensity applications. The acceleration process takes place continuously, and neither the RF frequency nor the magnetic bending field has to be cycled. The separation of subsequent turns allows continuous extraction of the beam from the cyclotron. Thus, the production of a continuous-wave (CW) beam is a natural feature of cyclotrons.

\section{Classification of Fixed Field Accelerators} \label{sec:classification}

The common property of the accelerator concepts discussed here is the fixed magnetic field, while in synchrotrons the field is varied during acceleration. Magnets have typically a significant inductance, corresponding to significant energy stored in the magnetic field. And as a result the cycling speed of magnets is limited. Rapid cycling synchrotrons can reach cycle frequencies of several 10\,Hz at maximum and the~operation of such machines is associated with a high power consumption. Most types of FFA require cycling of the RF frequency, and this can be done at rates of kHz.
The main challenge of fixed field accelerators is to provide sufficient focusing for the beam transport in both planes, and at the~same time to ensure synchronous acceleration. Historically the development started with the classical cyclotron that is simple to build but is limited by relativistic effects at proton kinetic energies of roughly 10\,MeV. Higher energies were reached by synchro-cyclotrons in pulsed operation and isochronous cyclotrons with more complicated fields. Relatively early also FFA concepts, the most general form of fixed field accelerators, were studied. The FFA employs strong focusing by alternating gradients, resulting in a~compression of orbits in a smaller radial space. The smaller radius variation leads to an overall compact and cost efficient accelerator design. Another advantage of FFA is a large acceptance in all planes. For all concepts beam injection and extraction is a critical aspect that becomes even a dominating problem for Megawatt class intensities. Starting from the classical cyclotron we will cover here the basics of the~isochronous cyclotron, the synchro-cyclotron and the FFA concepts. The types are compared in Table \ref{types}.

\begin{table}[h]
    \caption{Comparison of fixed field accelerator types.}
    \label{types}
    \begin{tabular}{ | p{3.94cm} | p{5.35cm} | p{5.35cm} |}
     \hline\hline
& \textbf{Focusing concept} & \textbf{Synchronous acceleration} \\\midrule
Classical cyclotron 
& weak focusing by slightly negative field index $-1 < k < 0$        
& only synchronous for $\gamma \approx 1$, i.e. $E_k \lesssim 10\,$MeV \\\hline
Isochronous cyclotron\newline  (sector focused cyclotron, Thomas cyclotron)
& focusing by azimuthally varying field (Flutter) and spiral angle 
& isochronous through positive average field index $k\approx\gamma^2-1>0$  \\\hline
Synchro-cyclotron
& weak focusing by slightly negative field index $-1 < k < 0$ 
& frequency sweep required,\newline thus pulsed operation\\ \hline
FFA
& alternating gradient focusing,\newline smaller orbit radius variation
& in general frequency sweep required but also versions with fixed frequency (e.g. serpentine acceleration) under discussion\\
        \hline\hline
    \end{tabular}
\end{table}

\section{Phase Stability in the Classical Cyclotron and the Concept of the Synchro-Cyclotron }
\label{sec:synchro}

For illustration purposes we address first the acceleration process in a classical cyclotron, accepting a~variation of the circulation frequency while the RF frequency is fixed. In a second part the concept of the synchro-cyclotron is discussed, that foresees an RF frequency sweep to accelerate the particles resonantly to higher energy.

In a homogeneous magnetic field the cyclotron frequency of a particle reduces as $1/\gamma$ with increasing energy. It could be kept constant if the field would increase radially in proportion to $\gamma$. However, a~positive field index leads to a loss of vertical focusing as shown in Section \ref{sec:classical}. Following a numerical example in \cite{kleeven}, for a 10\,MeV proton cyclotron one could use a slightly negative field index of $k=-0.01$. This corresponds to a vertical tune of $\nu_z= \sqrt{-k}=0.1$, or one oscillation in 10 turns. Taking into account the variation in arrival time, the energy gain of the particles is given by $\Delta E_k = e V_g \sin \phi$, where $\phi$ is the phase of the particle wrt. the RF phase. During the acceleration process the phase change per turn can be calculated by 

\begin{align}
\Delta \phi &= 2\pi\,\frac{B(r)-B_\RM{iso}}{B_\RM{iso}} \, . 
\label{dphi}
\end{align} 

\begin{figure} 
\centering\includegraphics[width=0.62\textwidth]{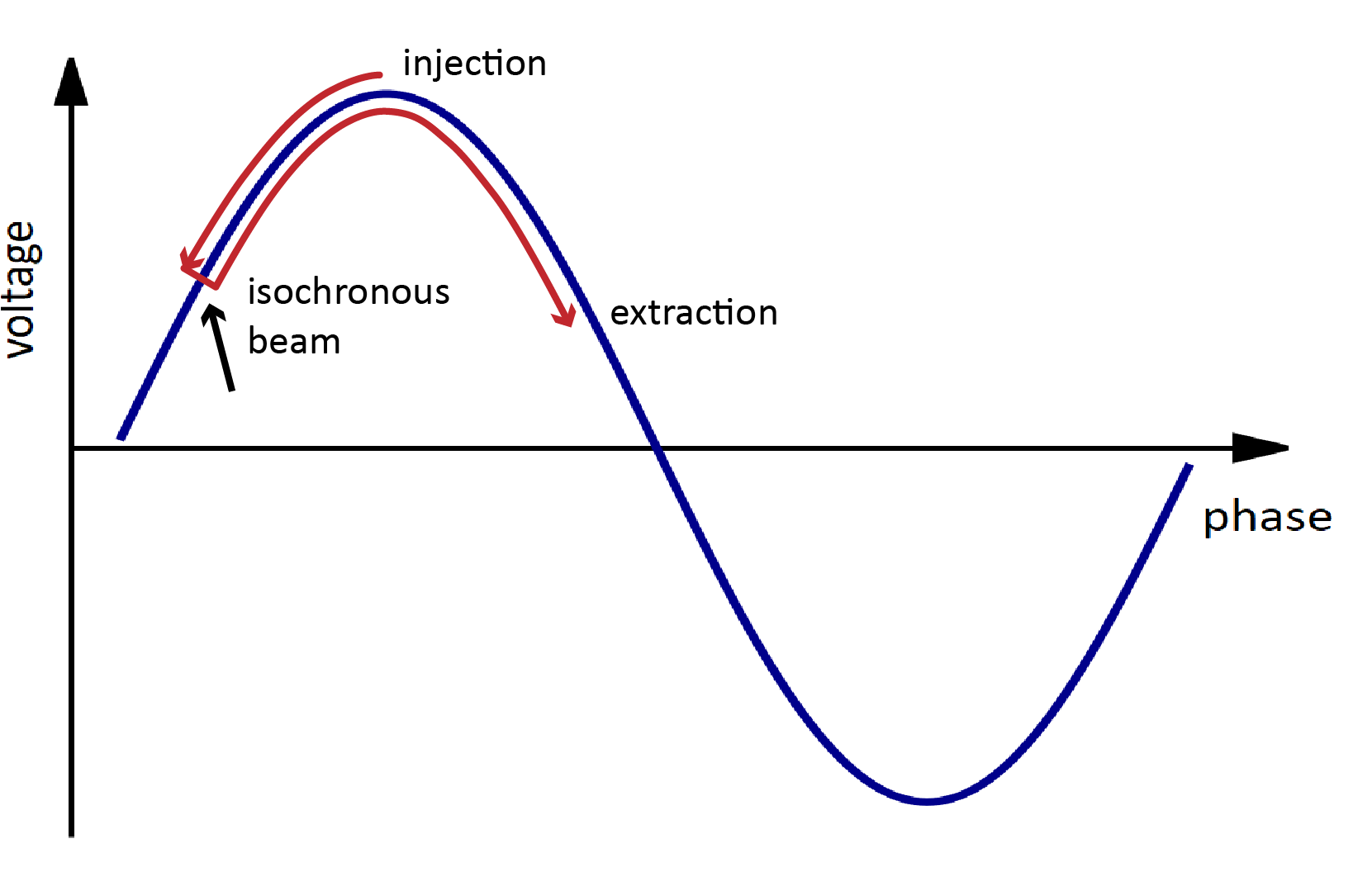}
\caption{Non-synchronous acceleration in a classical cyclotron after F.Chautard, GANIL.}
\label{fig:nonsynch}
\end{figure}

Here $B_\RM{iso}$ is the field that would result in a constant circulation frequency. In this situation one tries to keep the beam as long as possible in the upper half-wave of the sine function. The RF frequency is chosen in such a way as to obtain for the low energies a higher circulation frequency of the particles. During this part of the acceleration process the beam arrives earlier and is moving towards lower phase values (Fig.\,\ref{fig:nonsynch}). Since the voltage is still positive the energy is increasing from turn to turn and at some point circulation- and RF-frequency are matching, i.e. the beam is isochronous. As the acceleration proceeds the beam is now moving towards larger phase values. The acceleration process must finish before the phase slip reaches 180\,deg. Beyond this value the beam would be decelerated again.

\begin{figure} 
\centering\includegraphics[width=\textwidth]{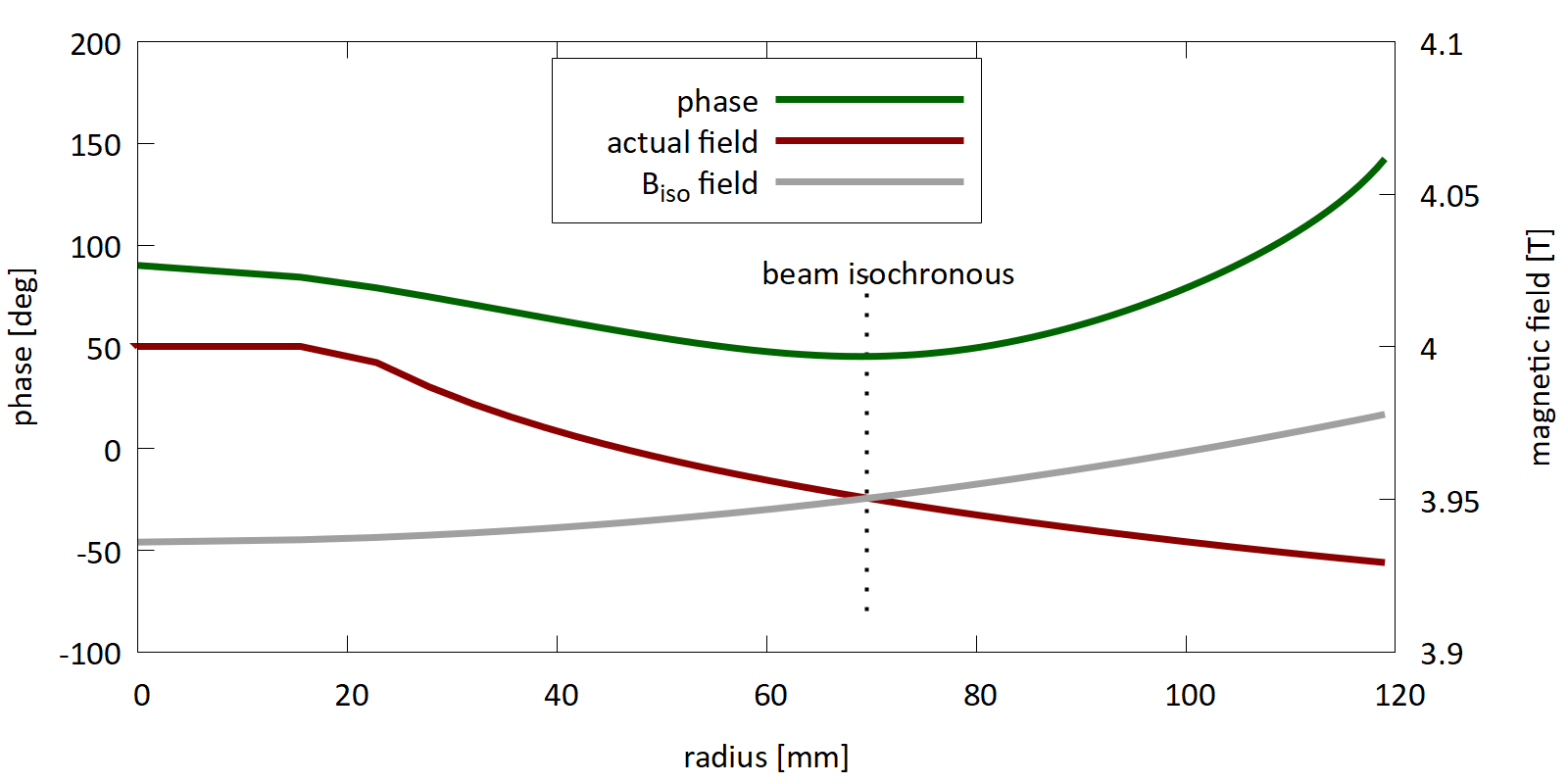}
\caption{Acceleration of protons in a 4 Tesla classical cyclotron to 10\,MeV. As a function of orbit radius are shown: the variation of the beams phase w.r.t. the RF accelerating voltage, the magnetic field with a negative slope to ensure focusing in both planes and the field $B_\RM{iso}$ that would result in a circulation frequency that matches the RF frequency.}
\label{fig:cycsimu}
\end{figure}

Figure\,\ref{fig:cycsimu} shows a numerical simulation of the described cyclotron. With a gap voltage of $2\times46\,$kV per turn and a magnetic field of 4\,T (s.c. magnet), a final energy of 10\,MeV is reached. This energy roughly presents the limit for acceleration in a classical cyclotron. The gap voltage needed for quick acceleration is  relatively high. The slippage of the beam phase results from two factors - the increase of the relativistic $\gamma$ factor and the decrease of the bending field with radius. In the described numerical example the total variation of revolution frequency is roughly 3\%. About $2/3$ of this change is caused by the field variation and $1/3$ by the relativistic mass increase. 
The kinetic proton energy of 10\,MeV, that is achieved in the example cyclotron, is sufficient for certain types of short lived isotope production as needed in hospitals. The advantage of the concept is a compact cyclotron with a weak-focusing rotational symmetric field.

In order to reach higher higher kinetic energies it must be ensured to keep the RF frequency synchronous with the circulation frequency of the beam. One way to achieve this is a modulation of the~RF frequency to adopt its time dependence to the beam circulation. This idea of the synchro-cyclotron was first proposed by Veksler \cite{veksler} and McMillan \cite{mcmillan} in papers in which they focused actually on the~synchrotron and its concept of phase stability.
A synchro-cyclotron  contains a simple, rotationally symmetric magnet with a slightly negative radial slope of the magnetic field. The absence of flutter focusing results in a higher average bending field as for isochronous cyclotrons (Section \ref{sec:avf}), and compact designs of synchro-cyclotrons are possible. The frequency of the RF system is modulated electronically or by a~mechanical system. Pulse rates of several hundred Hz up to kHz are possible. On the downside the~average beam intensity from a~synchro-cyclotron is at least two orders of magnitude lower than from a~cyclotron in CW operation. 
A technical key factor for a synchro-cyclotron is the correct modulation of the RF frequency. The required rate of change can be computed using the formula of the revolution frequency~(\ref{omega_rev}) and assuming a radially uniform field for now. Logarithmic differentiation gives:

\begin{equation}
\frac{\Delta \omega_\RM{rev}}{\omega_\RM{rev}} = - \frac{\Delta \gamma}{\gamma} \, .
\label{frate1}
\end{equation}

Using the fact that $\omega_\RM{rf} = h \omega_\RM{rev}$, and $\dot{\omega}_\RM{rf} \approx \Delta \omega_\RM{rf}/\tau$ with $\tau$ the revolution time, this can be converted into a relation for the ramp rate of the RF frequency:

\begin{equation}
\frac{e V_g \sin\phi_s}{E_k + m_0 c^2} = - \frac{2\pi}{\omega_\RM{rf}^2} \, \dot{\omega}_\RM{rf}
\, .
\label{frate2}
\end{equation}

A representative example of a modern synchro-cyclotron is the superconducting S2C2 machine of IBA \cite{kleeven2}. This cyclotron with a $5.7\,$T magnet is part of the compact ProtheusOne cancer treatment system, suited for hospitals. For the frequency modulation the system uses fast rotating electrodes (ROTCO) to achieve a relatively high pulse repetition rate of 1\,kHz. The beam current is in the range of nA, which is lower than the intensity obtained with isochronous cyclotrons, but which is adequate for cancer therapy.

\section{The isochronous Cyclotron} \label{sec:avf}

Another way to overcome the problem of insufficient vertical focusing is the introduction of azimuthally varying fields, as it is realised in the Thomas-, or AVF-cyclotron. The principle was proposed in 1938 by L.H.\,Thomas~\cite{thomas}, but it took several decades before a cyclotron based on this principle was actually built (in Delft in 1958). The variation of the vertical bending field along the trajectory leads to transverse forces on the particles that provide suitable focusing characteristics in both of the transverse planes. In a Thomas cyclotron, the average field strength can be increased as a function of radius without losing the vertical stability. As a result, this concept allows higher kinetic energies to be achieved, for example 1~GeV for protons. The required field variation in the cyclotron can be realised by special shaping of the poles of a compact single magnet. The focusing can be further increased by the introduction of spiral sector shapes, and sector boundaries that have tilt angles with respect to the cyclotron center. With such magnet configurations, the squared betatron frequencies are approximately:

\begin{eqnarray}
\nu_r^2 & = & 1+k \nonumber \\
\nu_z^2 & = & -k + F^2 (1+2\tan^2 \alpha). 
\label{cyc_tunes}
\end{eqnarray}

\begin{figure}[h!] 
\centering\includegraphics[width=0.65\textwidth]{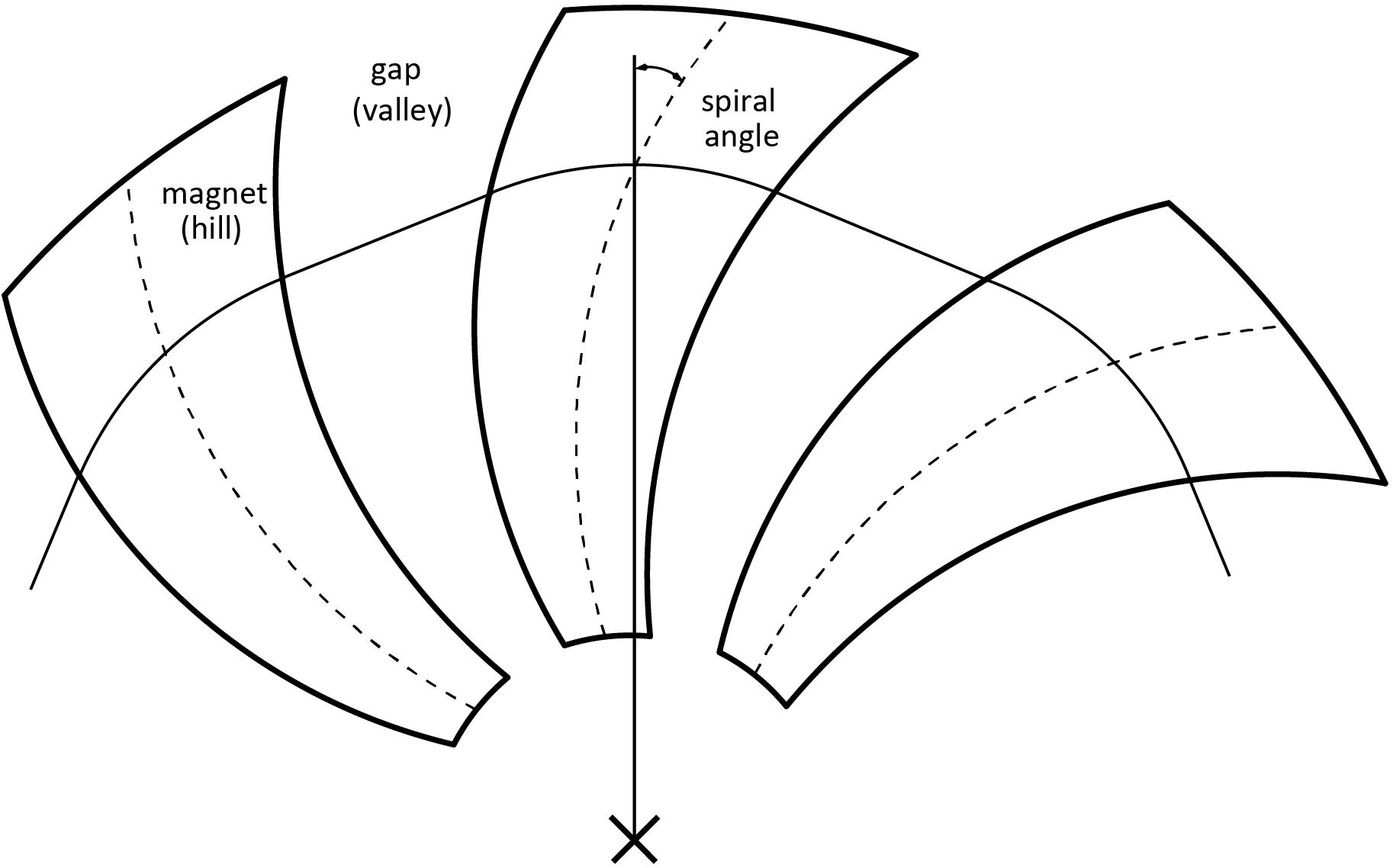}
\caption{Spiral magnet sectors in an azimuthally varying field (AVF) cyclotron and definition of the average spiral angle.} 
\label{fig:spiral_angle} \end{figure}

The so-called flutter factor $F$ equals the relative root mean square (r.m.s.) variation of the bending field around the circumference of the cyclotron. The spiral angle $\alpha$ is defined as shown in Fig.~\ref{fig:spiral_angle}.

\begin{equation}
F^2 = \frac{\left< B_z^2 \right> -\left< B_z \right>^2}{\left< B_z \right>^2}.
\label{flutter}
\end{equation}

The next and most recent step in the history of cyclotron development was the introduction of separated-sector cyclotrons. Such cyclotrons have a modular structure, consisting of several sector-shaped dipole magnets and RF resonators for acceleration. The modular concept makes it possible to construct larger cyclotrons that accommodate the bending radii of ions at higher energies. As compared to a synchro-cyclotron, the sector-focused cyclotron is an attractive solution for high-intensity applications, owing to its advantage of CW operation. 

For clean extraction with a septum, the distance between the turns at the extraction radius must be maximized. It is therefore essential in the design of a high-intensity cyclotron to consider the transverse separation of the beam. To calculate the step width per turn, we start from the formula for the (average) magnetic rigidity,

\begin{equation} BR = \frac{p}{e} = \sqrt{\gamma^2-1} ~\frac{m_0c}{e}.
\end{equation}

By computing the total logarithmic differential, we obtain a relation between the changes in  radius, magnetic field and energy of the particle beam:

\begin{equation} 
\frac{\RM{d}B}{B} +\frac{\RM{d}R}{R} = \frac{\gamma~\RM{d}\gamma}{\gamma^2-1}.
\end{equation}

Using the field index $k$ given in Eq. (\ref{index}), we obtain

\begin{equation} 
1+k = \frac{\gamma R}{\gamma^2 -1} \frac{\RM{d}\gamma}{\RM{d}R}.
\nonumber \end{equation}

Noting that the change in the relativistic quantity $\gamma$ per turn is $\RM{d}\gamma/\RM{d}n_\RM{t} = U_\RM{t}/(m_0c^2)$, where $U_\RM{t}$ denotes the energy gain per turn, we finally obtain the radius step width:

\begin{eqnarray}
\frac{\RM{d}R}{\RM{d}n_\RM{t}} & = & \frac{\RM{d}\gamma}{\RM{d}n_\RM{t}} \frac{\RM{d}R}{\RM{d}\gamma} \nonumber \\
& = & \frac{U_\RM{t}}{m_0c^2} \frac{\gamma R}{(\gamma^2-1)(1+k)} \label{step_1} \\
& = & \frac{U_\RM{t}}{m_0c^2} \frac{\gamma
R}{(\gamma^2-1)\nu_r^2}. \label{step_2}
\end{eqnarray}

In the outer region of the cyclotron, near the extraction radius, it is possible to violate the condition of isochronicity for a few turns. By reducing the slope of the field strength, which is related to the radial tune, it is possible to increase the turn separation locally. In the~fringe field region of the magnets, the~field decreases naturally. By going from Eq. (\ref{step_1}) to Eq. (\ref{step_2}) using Eq. (\ref{nu_r}), we can show the~relation between the step width and the radial tune. If the condition of isochronicity remains valid, the~dependence on the~field index and the radial tune can be eliminated, and the step width is given by

\begin{equation}
\frac{\RM{d}R}{\RM{d}n_\RM{t}}  =
\frac{U_\RM{t}}{m_0c^2} \frac{R}{(\gamma^2-1)\gamma}.
\label{step_full}
\end{equation}

In this form, the equation shows the strong dependence of the step width on the beam energy. Above 1~GeV, it becomes very difficult to achieve clean extraction with an extraction septum. In the~non-relativistic limit $\beta \ll 1$ we can use $\gamma (\gamma^2-1) \approx \beta^2$, $\beta^2 = R^2/R_\infty^2$ to show the scaling with the~inverse of the radius, as it was indicated in Fig.\,\ref{fig:spiral}:

\begin{equation}
\frac{\RM{d}R}{\RM{d}n_\RM{t}}  =
\frac{U_\RM{t}}{m_0c^2} \, \frac{R_\infty^2}{R}, \,\RM{for}\, \beta \ll 1.
\label{step_classic}
\end{equation}

\begin{figure}[h!] 
\centering\includegraphics[width=0.75\textwidth]{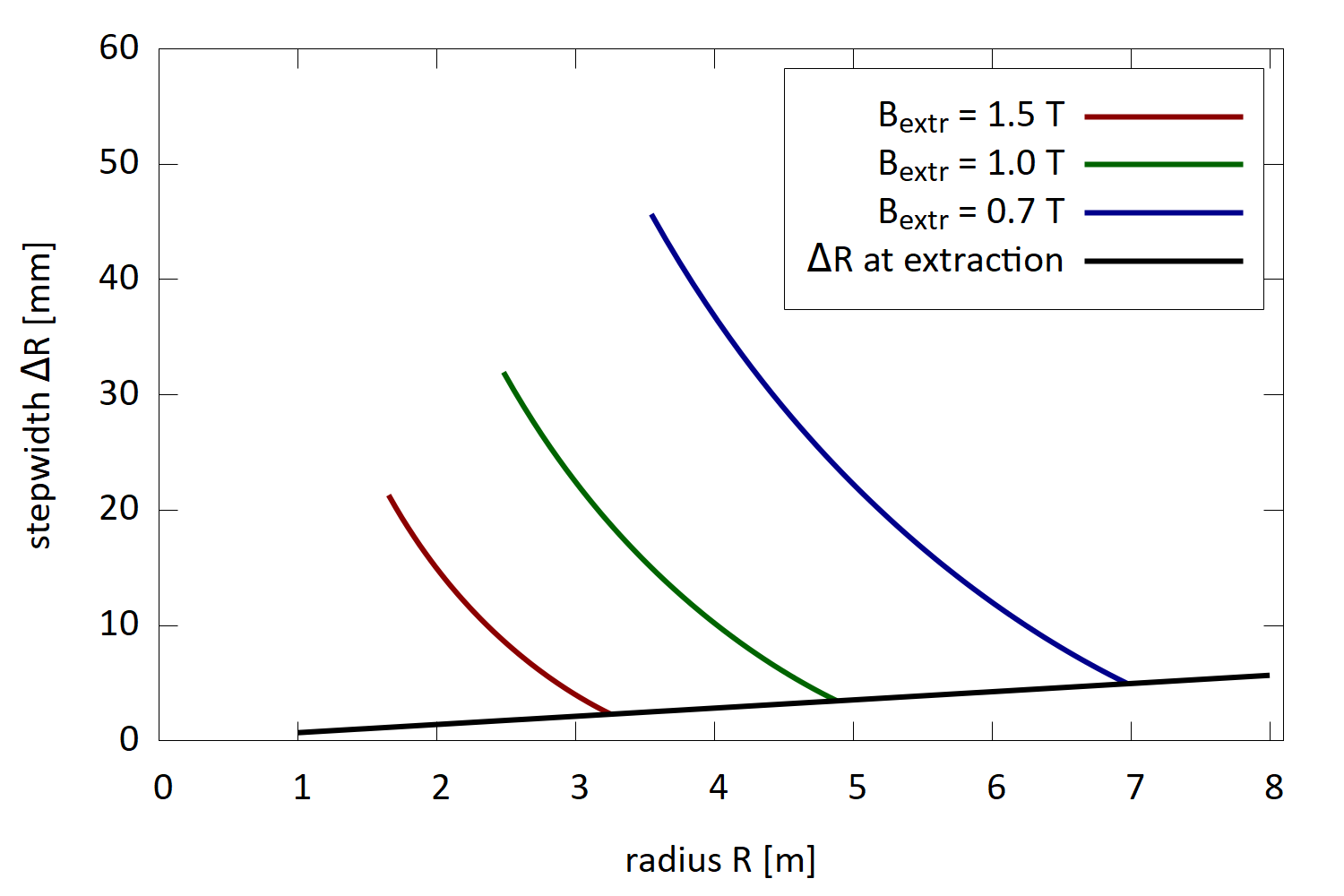}
\caption{Behaviour of radial turn separation during acceleration from 100\,MeV to 800\,MeV for three cyclotrons with varying size and field strength. The final turn separation at extraction scales with the size.} 
\label{fig:turnsep} \end{figure}

In Fig.\,\ref{fig:turnsep} the relevance of Eq. (\ref{step_full}) is demonstrated with the numerical example of three field strengths for cyclotrons that accelerate a beam from $100\,$MeV to $800\,$MeV. While for every case the step width decreases during the course of acceleration, the important turn separation at extraction becomes larger with increasing extraction radius and overall size of the cyclotron. 

An effective way to increase the turn separation at the extraction element is the introduction of orbit oscillations by deliberately injecting the beam slightly off centre. When the phase and amplitude of the orbit oscillation are chosen appropriately, and also the behaviour of the radial tune is controlled in a~suitable way, the beam separation can be increased by a factor of three. According to Eq. (\ref{step_1}), this gain is equivalent to a cyclotron three times larger and is thus significant. This scheme is used in the~PSI Ring cyclotron. In Ref. \cite{bi}, the beam profile in the outer turns was computed numerically for realistic conditions, and the results are in good agreement with measurements.

In summary, the clean extraction of the beam is of utmost importance for high-intensity cyclotron operation. The turn separation at the extraction element can be maximized by the following measures.

\begin{itemize}
\item The extraction radius should be large, i.e., the overall dimensions of the cyclotron should be chosen as large as reasonably possible.
\item The energy gain per turn should be maximized by installing a sufficient number of resonators with high performance. 
\item At relativistic energies, the turn separation diminishes quickly, and thus the final energy should be kept below approximately 1\,GeV.
\item In the extraction region, the turn separation can be increased by lowering the slope of the field index and by utilizing orbit oscillations resulting from controlled off-centre injection.
\end{itemize}

An alternative to the extraction method described here is extraction via charge exchange, for example several facilities accelerate H$^-$ ions to extract protons by stripping the electrons. More details of this method are given in Section \ref{sec:design} and Ref. \cite{seidel3}. Also the acceleration of H$_2^+$ has been considered for high intensity beams since this molecule is more stable \cite{calabretta}.

Besides the large turn separation, another ingredient of clean extraction are low beam tails. One important mechanism for tail production is longitudinal space charge, introducing energy spread, which is then converted into transverse tails. Tails produce losses, and in practice the losses must be limited to the order of $\approx100\,$W. Fast acceleration is the best method to reduce these tails. Joho has shown in Ref. \cite{joho} that the attainable intensity in the PSI Ring cyclotron scales as the inverse of the cubed number of turns, under the condition of roughly constant absolute losses. Indeed, over the history of the accelerator significant improvements in intensity were achieved by raising the accelerating voltage, and the achieved intensity is in agreement with Joho's scaling law.

Also transverse space charge forces present limitations for the bunch intensities. The vertical focusing is reduced by space charge and obviously an intensity limit is reached when the focusing term vanishes. This condition was used by Blosser \cite{handbook} to formulate a space charge limit for cyclotrons. Since individual turns in a cyclotron are close also the effect on neighbouring turns must be considered \cite{jianjun}.

\section{Design aspects of separated-sector cyclotrons} \label{sec:design}

\begin{figure} 
\centering\includegraphics[width=0.95\textwidth]{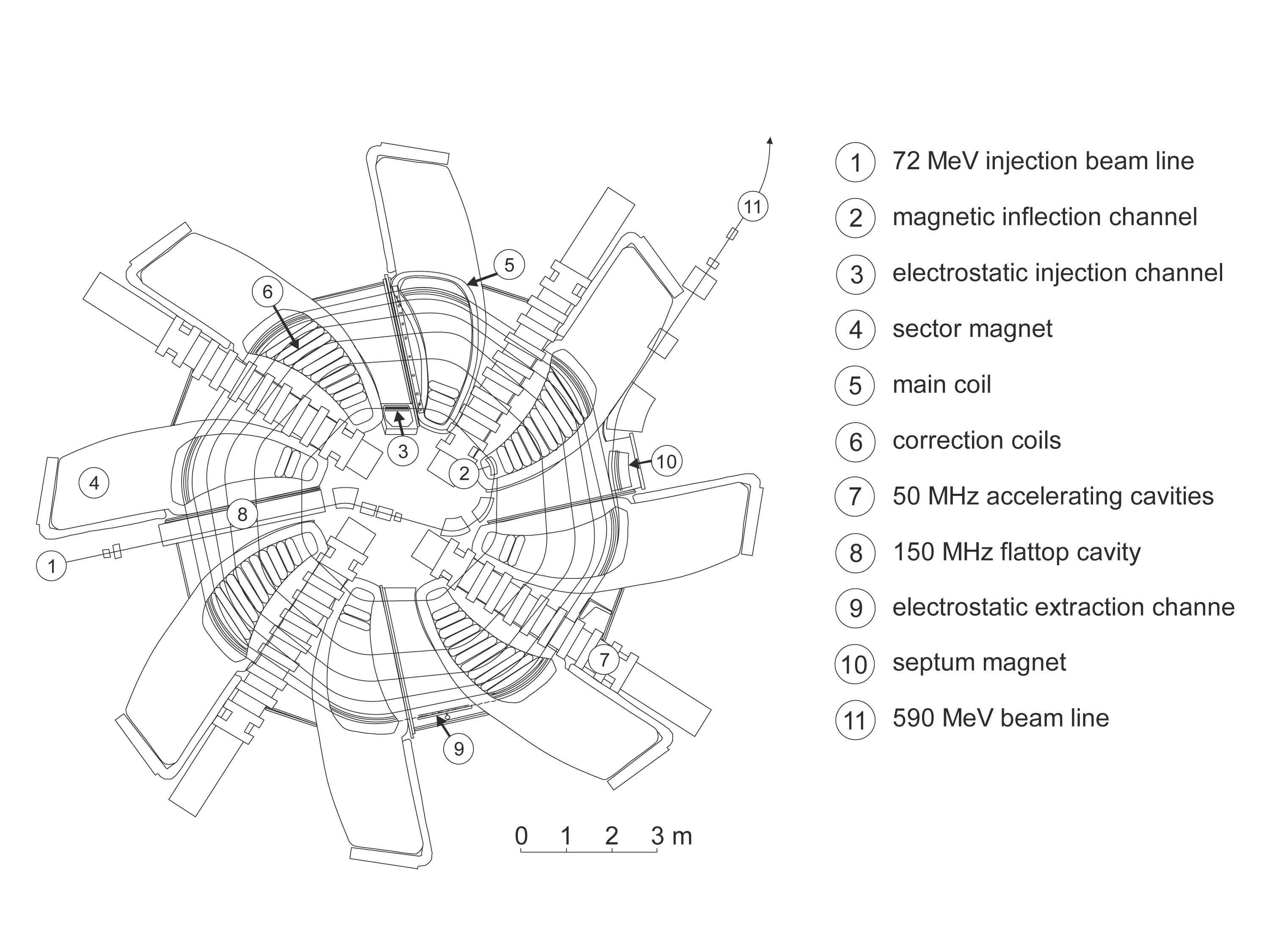}
\caption{Top view of the PSI Ring cyclotron. This separated-sector cyclotron contains eight sector magnets, four accelerating resonators (50~MHz), and one flat-top resonator (150~MHz).}
\label{fig:ring}
\end{figure}

Modern cyclotrons that are able to reach higher $K$-values and/or high beam intensity, are typically realised as separated-sector cyclotrons. They employ a modular concept involving a combination of sector-shaped magnets, RF resonators, and empty sector gaps to form a closed circular accelerator. The modular concept simplifies the construction of cyclotrons with diameters significantly larger than those achieved with the classical single-magnet concept. The large orbit radius at maximum energy permits extraction with extremely low losses. The modularity also has significant advantages concerning the serviceability of the accelerator, especially in view of the need to handle activated components. A simplified view of the PSI Ring cyclotron is given as an example in Fig.~\ref{fig:ring}.

During the course of acceleration, the revolution time is kept constant, leading to a significant variation in the average orbit radius. The lateral width of the elements in the ring is large in comparison with the elements of a synchrotron that uses strong focusing. The mechanical design of the vacuum chambers and sealed interconnections is thus challenging. On the other hand the large variation in the~radius makes it possible to separate the turns at the outer radius and to realize an extraction scheme for CW operation with very low losses. In comparison the close orbit spacing in FFA rings reduces cost but makes continuous extraction difficult. The extraction loss is the limiting effect for high-intensity operation of cyclotrons. The wide vacuum chambers (2.5~m for the PSI Ring cyclotron) require special sealing techniques. In a cyclotron, as in a single-pass accelerator, vacuum levels of $10^{-6}$~mbar are sufficient for the acceleration of protons. So-called inflatable seals are manufactured from thin steel sheets with two sealing surfaces per side and an intermittently evacuated volume between. To simplify installation, these seals are positioned on radial rails between two elements. Inflation with pressurized air seals the~surfaces. This screwless scheme can tolerate small positioning errors and has the advantage of short mounting times in a radiation environment.

Cyclotron magnets are built as single magnets for compact cyclotrons that achieve AVF properties through the pole shape. Modern compact cyclotrons use a superconducting magnet to achieve higher field strength with existing examples up to 9\,Tesla. For sector cyclotrons separate magnets are used, often with spiral shapes. The mechanical design is challenging if a wide radius variation of the beam has to be covered. 

\begin{figure} 
\centering\includegraphics[width=0.50\textwidth]{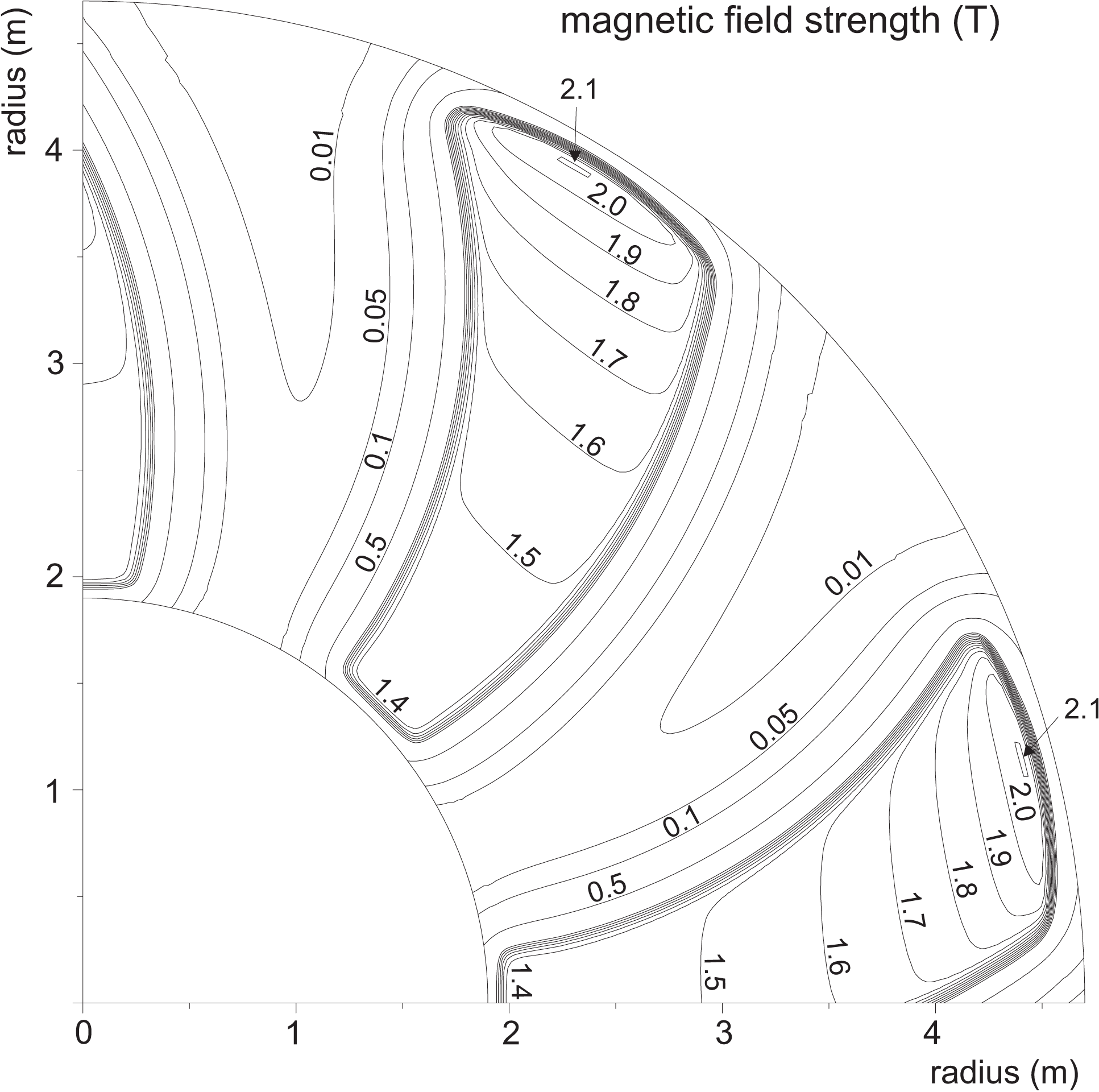}~ \includegraphics[width=0.49\textwidth]{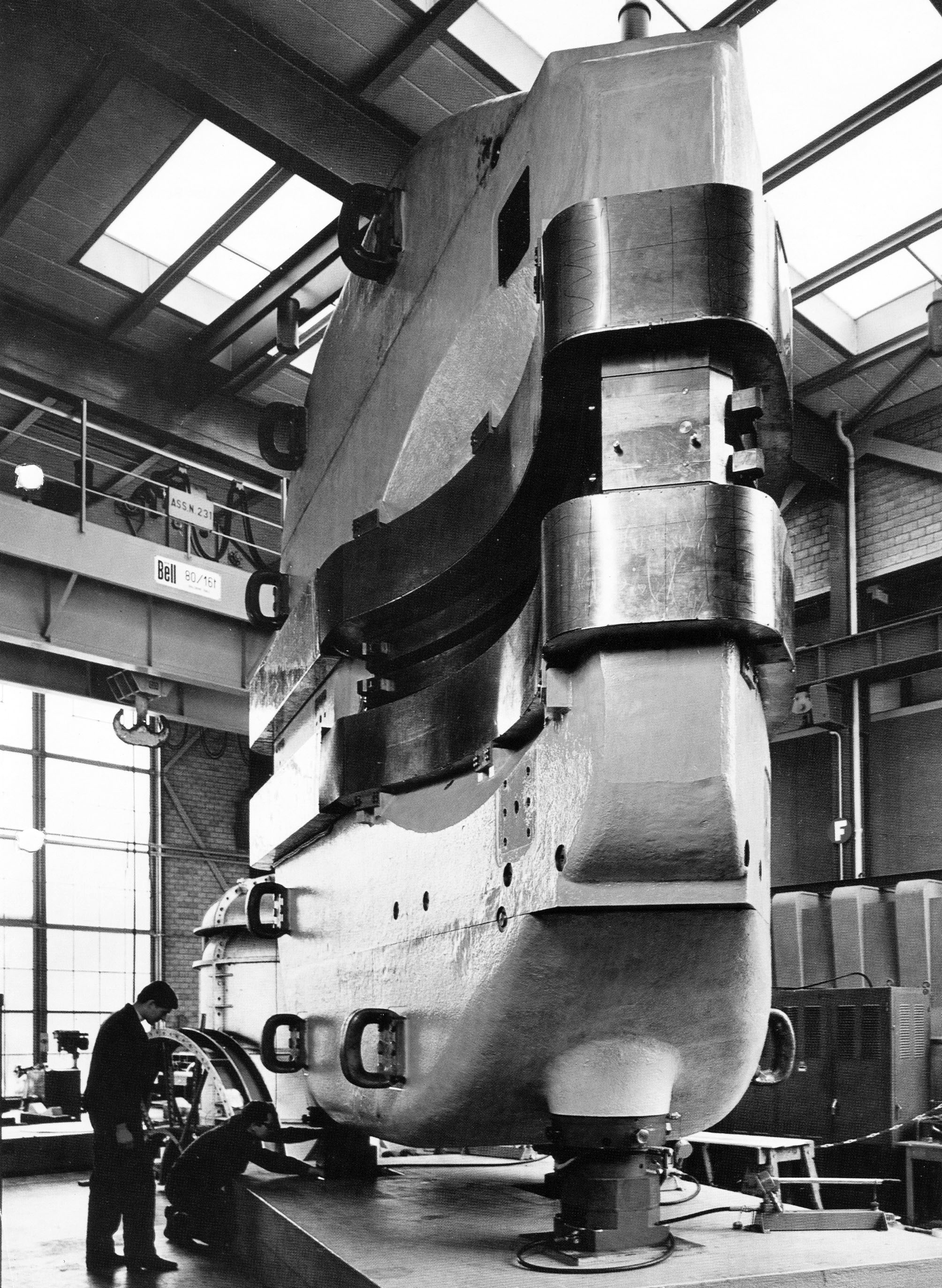}
\caption{Left: fieldmap of the PSI Ring cyclotron sector magnets showing increasing field strength with radius. Right: photograph of the sector magnet before installation.}
\label{fig:magnet} \end{figure}

\begin{figure} 
\centering\includegraphics[width=0.5\textwidth]{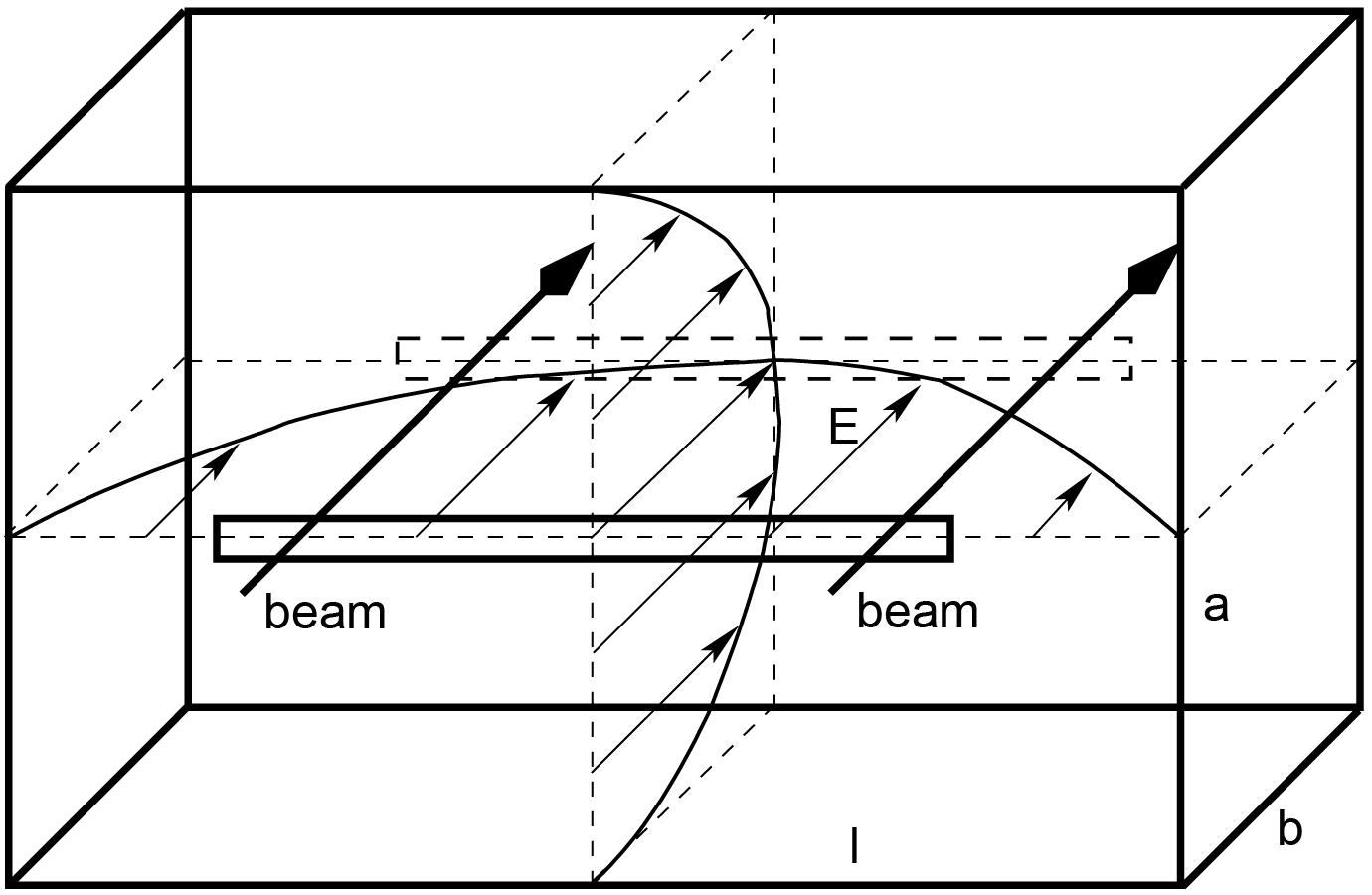}
\caption{Field distribution and orientation of the beam in a cyclotron box resonator} 
\label{fig:boxcav} \end{figure}

\begin{figure} 
\centering\includegraphics[width=0.92\textwidth]{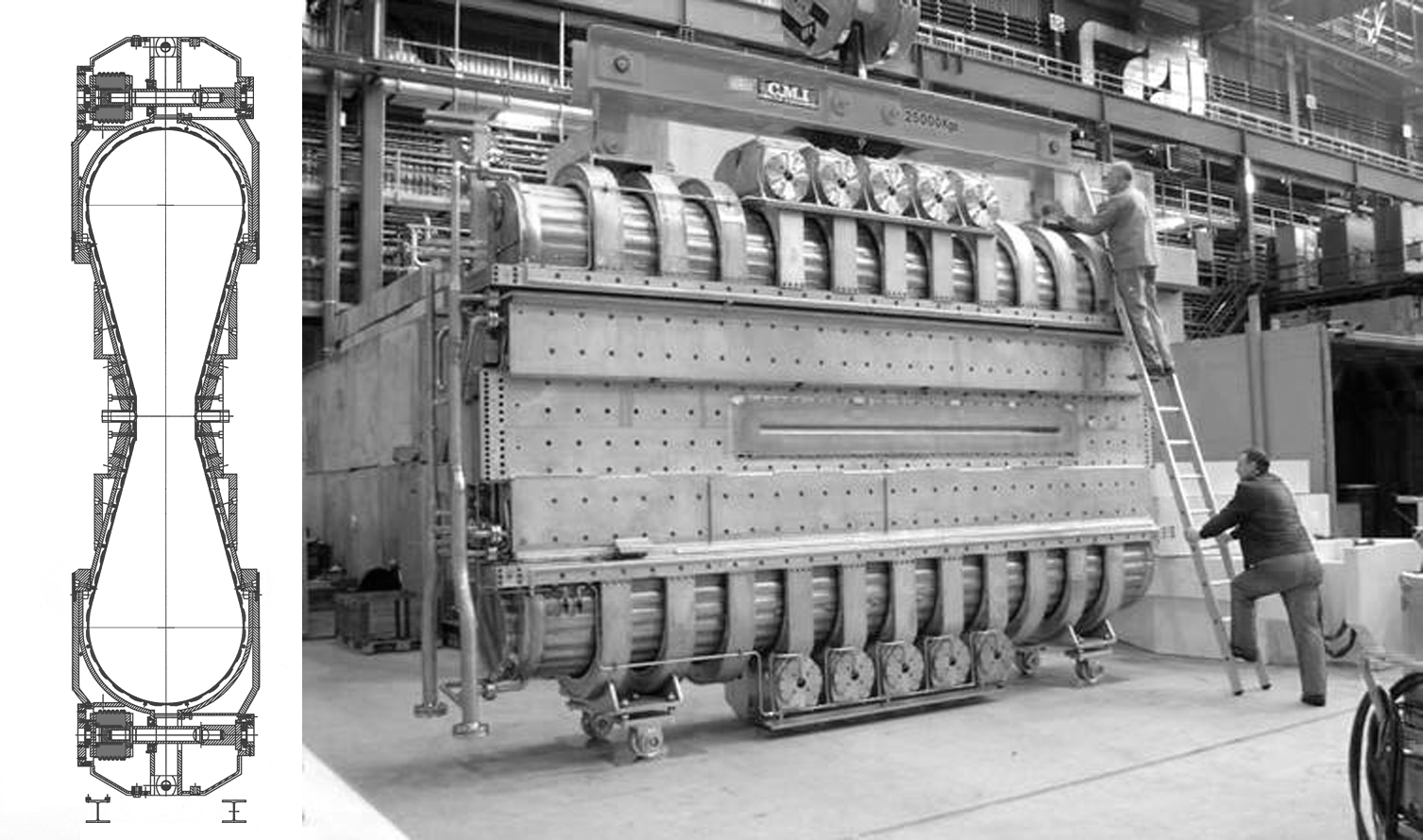}
\caption{Cross-section of PSI resonator (left), and photograph
(right)} \label{fig:cav_picture} \end{figure}

In compact cyclotrons the RF acceleration is still provided by Dee-shaped electrodes, but often four electrodes are used for more efficient acceleration. In separated sector cyclotrons one uses box resonators in a similar arrangement as for synchrotrons. However, for sector cyclotrons the resonators have a more rectangular shape, wide enough to cover the range of orbit variation. The beam passes through a slit. The field distribution in such a resonator is shown in Fig.\,\ref{fig:boxcav}. The electric field in the plane of the beam varies with the radial coordinate as a sine function. In this configuration, the resonance frequency depends on the radial length $l$ and the height $a$ of the cavity:
\begin{equation}
f_0 = \frac{c}{2} \sqrt{\frac{1}{a^2} + \frac{1}{l^2}} .
\end{equation}
Thus the frequency is independent of the azimuthal width $b$ of the cavity. In practice, the shape of the~cavity is not made exactly rectangular. Instead, the azimuthal width is reduced in the midplane (Fig.~\ref{fig:cav_picture}) to minimize the travel time of the particles in the field. In the case of the PSI cyclotrons, all accelerating resonators are operated at 50.6~MHz. In the Ring cyclotron, the resonators are made from copper and achieve a quality factor of $4.8\times 10^4$. The typical gap voltage is 830~kV. Each resonator can transfer 400~kW of power to the beam.

The concept of the separated-sector cyclotron requires external injection of a beam of good quality. Both injection and extraction are often performed using an electrostatic deflection channel. The beam is deflected at a certain radius, while the neighbouring turns must not be affected. This is achieved by placing a thin electrode between the two turns. Particles in the beam tails that hit this electrode are scattered, and these generate losses and activation. A magnetic element would need much more material to be placed between the turns.  

\begin{figure} 
\centering\includegraphics[width=0.95\textwidth]{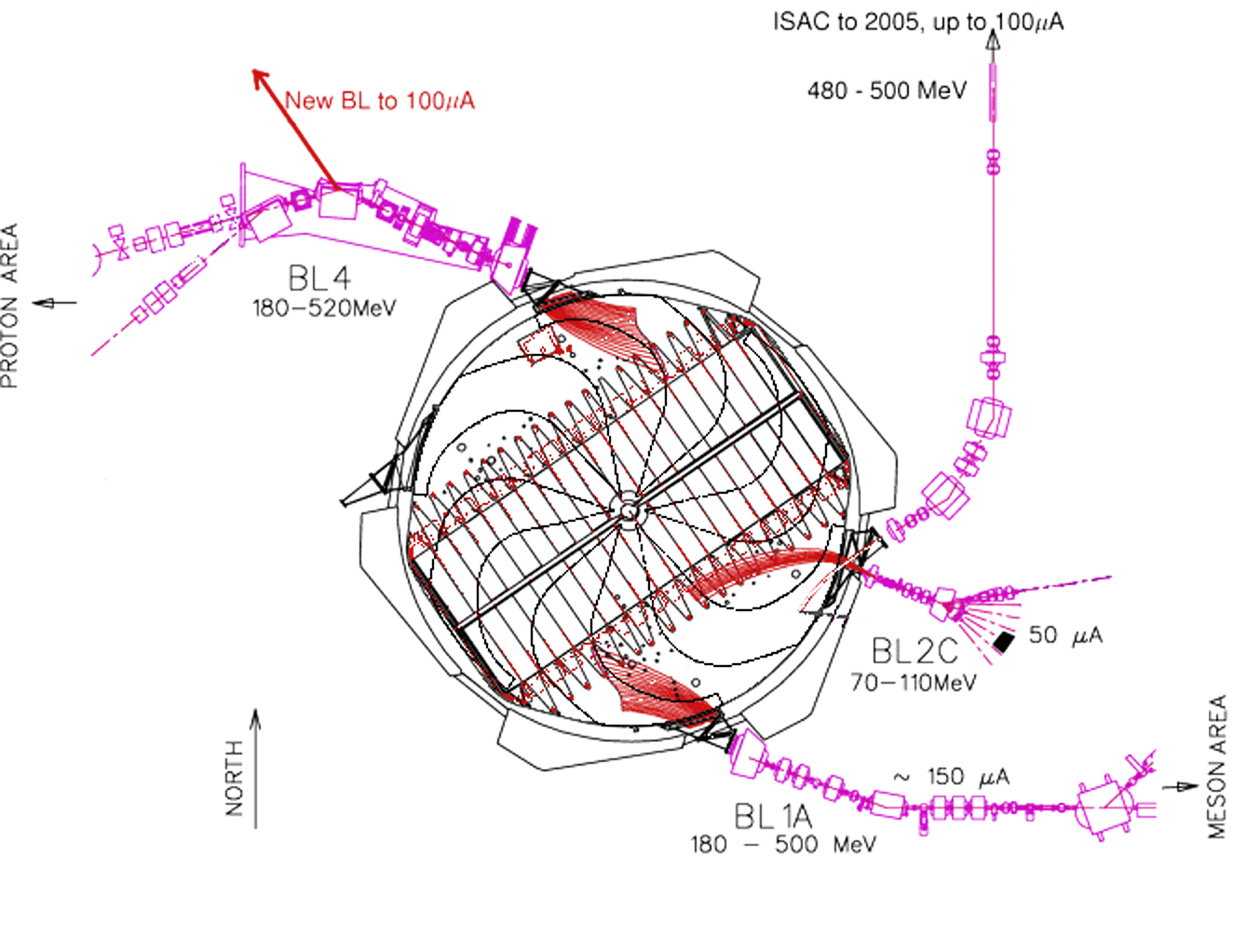}
\caption{Acceleration of H$^-$ ions in the TRIUMF cyclotron and stripping extraction allows to provide multiple beams in parallel at variable energy (courtesy TRIUMF).}
\label{fig:triumf}
\end{figure}

Stripping or charge exchange extraction is another elegant scheme to extract a beam from a cyclotron. In this case ions are accelerated that are not fully ionized. The electrons are removed from the~ions by passing them through a thin foil. The sudden change of the charge to mass ratio causes a change of the curvature of the ion path in a magnetic field. In this way the stripped ions are separated from the circulating beam and they are easily extracted. A prominent example for this technology is the acceleration of H$^-$, a proton with two bound electrons. After removing the two electrons the orbit curvature is inverted. By introducing the stripping foil at varying radius it is even possible to extract beam at different energies, Fig.\,\ref{fig:triumf}.
While the advantages of the method are obvious, a number of difficulties must be mentioned. The second electron of H$^-$ is bound rather weakly to the proton. In a magnetic field or by interaction with residual gas atoms, the ion is dissociated with relatively high probability. For example in the TRIUMF cyclotron the maximum magnetic field is limited to about $0.6\,$T for this reason. 

Some parameters of large cyclotrons operating today are listed in Table~\ref{tab:cyclotrons}. The TRIUMF cyclotron~\cite{dutto} accelerates H$^-$ ions and utilizes the mentioned stripping extraction. The RIKEN Ring cyclotron~\cite{kase} is not a high-intensity machine, but it allows a broad variety of ions to be accelerated. It employs superconducting sector magnets \cite{okuno}, which deliver a very high bending strength, reflected by the corresponding $K$-value. The PSI Ring cyclotron was proposed in the 1960s by Willax \cite{willax}. It is specialized for high-intensity operation at the expense of reduced flexibility \cite{seidel}.

\begin{table}[h!] \centering 
\caption{Selected parameters of large sector cyclotrons. The TRIUMF cyclotron uses a single magnet with sector poles.}
\label{tab:cyclotrons}
\begin{tabular}{|@{\hspace*{1.5mm}}c@{\hspace*{1.5mm}}|c|c|c|c|c|c|@{\hspace*{1.2mm}}c@{\hspace*{1.2mm}}|c|}
\hline\hline
\rule{0mm}{6mm} \textbf{Cyclotron} & ${\pmb K}$ & ${\pmb N_\RM{mag}}$ & ${\pmb h}$ & ${\pmb R_\RM{inj}}$ & ${\pmb R_\RM{extr}}$ & \textbf{Extraction} & \textbf{Overall} & \textbf{Application}\\
\rule[-3mm]{0mm}{6mm} & \textbf{[MeV]} & & & \textbf{[m]} & \textbf{[m]} & \textbf{method} & \textbf{transmission} & \\
\hline \rule{0mm}{6mm} 
TRIUMF & 520 & 6 & 5 & 0.25 & 3.8--7.9 & H$^-$ stripping & 0.70 & variable energy, \\
& & (sect.) & & & & channel & & multiple beams\\
\hline \rule{0mm}{6mm} 
PSI Ring & 592 & 8 & 6 & 2.1 & 4.5 & Electrostatic & 0.9998 & high intensity\\
& & & & & & channel & & \\
\hline \rule{0mm}{6mm} 
RIKEN & 2600 & 6 & 6 & 3.6 & 5.4 & Electrostatic & (varies) & variable ions\\
\rule[-3mm]{0mm}{6mm} Ring & & & & & & channel & & \\
\hline\hline
\end{tabular}
\end{table}

\section{Fixed Field alternating Gradient Accelerator} \label{sec:ffa}

The FFA is the most generic concept among the fixed field accelerators and it was proposed following the discovery of alternating gradient focusing in the~1950s. In FFA alternating gradient focusing is the~dominating focusing effect \cite{craddock}, although also edge focusing is relevant. Several groups in the US, in Japan and in the USSR worked independently on FFA concepts. A well known effort was undertaken by the Mid Western Universities Research Association (MURA) with the prominent members Symon and Kerst \cite{symon}. MURA has built and tested electron models of FFA. Later FFA were built for protons at KEK and KURRI (Kyoto University). With improved RF and magnet technology the interest in FFA concepts has grown significantly over the last decade. A fast acceleration process is important to avoid beam losses through resonances and nonlinear field components. Some concepts require unusual field profiles that could be realised in superconducting magnets. Since the magnets are fixed in field and also the phase space acceptance of FFA is large, these machines have the potential for accelerating high intensity beams. Also the application for the acceleration of muon beams is discussed for the reason of large acceptance.

With compact magnets, particularly permanent magnets, a high number of quadrupoles per length can be realised. With such FFA focusing channels, even in a bending section the dispersion function can be suppressed to small values. A potential application are proton therapy gantries with large energy bandwidth \cite{gantry}.

\subsection{Scaling FFA}
In a fixed field accelerator the beam is moving on changing orbits, and also the focusing characteristics may be varying and results in tune changes. When resonances are crossed the beam may blow up and particles are lost. To avoid such variations the idea of a scaling FFA was developed. It involves an~energy independent focusing characteristics and self-similar, "scaled" orbit shapes.  
Using the field index averaged around the circumference, also for an FFA the transverse tunes are given by Eq. (\ref{cyc_tunes}). If the~tunes shall be independent of radius and beam energy, the field index $k$ must be kept constant. This is achieved by fields of the form:

\begin{equation}
B(r, \theta) = B_0 \left( \frac{r}{r_0} \right)^k f \left( \theta - \alpha \ln \left( \frac{r}{r_0} \right) \right)
\label{scaling}
\end{equation}

Note that contrary to isochronous cyclotrons with variable field index Eq. (\ref{index1}), for scaling FFA $k$ shall be constant over the range of radial orbit variation. For $\alpha=0$ we have a sector type field, otherwise a spiral field that enhances the vertical focusing. Equation (\ref{scaling}) also implies that $p \propto r^{k+1}$. Besides of constant tune the orbits at varying energy in a scaling FFA have a self-similar shape. 

A momentum compaction factor can be derived from the variation of the circumference path length~$C$:

\begin{align}
\alpha_c & \equiv \frac{\Delta C/C}{\Delta p/p} = \frac{1}{R} \avg{D_r} \nonumber \\[5pt]
& = \frac{1}{1+k}.
\label{mcompact}
\end{align}

Similarly the slip factor, relating momentum and circulation time, is given by:

\begin{align}
\eta_c & \equiv \frac{\Delta \tau/\tau}{\Delta p/p} = \alpha_c - \frac{1}{\gamma^2} \nonumber \\
& = \beta^2 - \frac{k}{1+k}.
\label{slipf}
\end{align}

Like in a synchrotron the slip factor changes its sign when the transition energy is passed. At the~transition energy the revolution frequency becomes independent of particle energy and we derive:

\begin{align}
\gamma_\RM{tr} &= \frac{1}{\sqrt{\alpha_c}} = \sqrt{1+k}.  
\label{gammat}
\end{align}

A sign change can occur for a positive field index. With negative field index the accelerator operates below transition. Using the slip factor a relation between energy change and frequency change can be established.

\begin{align}
\frac{d \omega_\RM{rev}}{\omega_\RM{rev}} &= - \frac{d\tau}{\tau} = -\eta_c \frac{dp}{p} \nonumber \\
&= -\frac{\eta_c}{\beta^2} \, \frac{dE}{E}. 
\label{slipd}
\end{align}

In principle Eq. (\ref{slipd}) can be integrated to obtain the required evolution of the revolution frequency and thus the RF frequency for a scaling FFA during the acceleration process. However, also two known relations can be used to obtain the same result. A first equation is derived from the magnetic rigidity:

\begin{align}
\omega_\RM{rev} &= \frac{\omega_c}{\gamma} \left( \frac{R}{R_0} \right)^k .
\label{omega1}
\end{align}

A second equation is obtained from the relation between speed of the particle and circulation frequency:

\begin{align}
\omega_\RM{rev} &= \frac{\beta c}{R}. 
\label{omega2}
\end{align}

We can now eliminate $R$ and express the angular frequency as an exclusive function of $\gamma$:

\begin{equation}
\omega_\RM{rev}(\gamma) = \omega_0 \, \frac{\gamma_0}{\gamma} 
\left( \frac{\gamma^2-1}{\gamma_0^2-1} \right)^
\frac{k}{2(k+1)}.
\label{omegaffa}
\end{equation}

Here a reference energy $\gamma_0 m_0 c^2$ is chosen, corresponding to a reference angular frequency $\omega_0$ which occurs at the radius $R_0$:

\begin{align}
\omega_0 &= \frac{\omega_c}{\gamma_0} .
\label{omega0}
\end{align}

\begin{figure} 
\centering\includegraphics[width=0.8\textwidth]{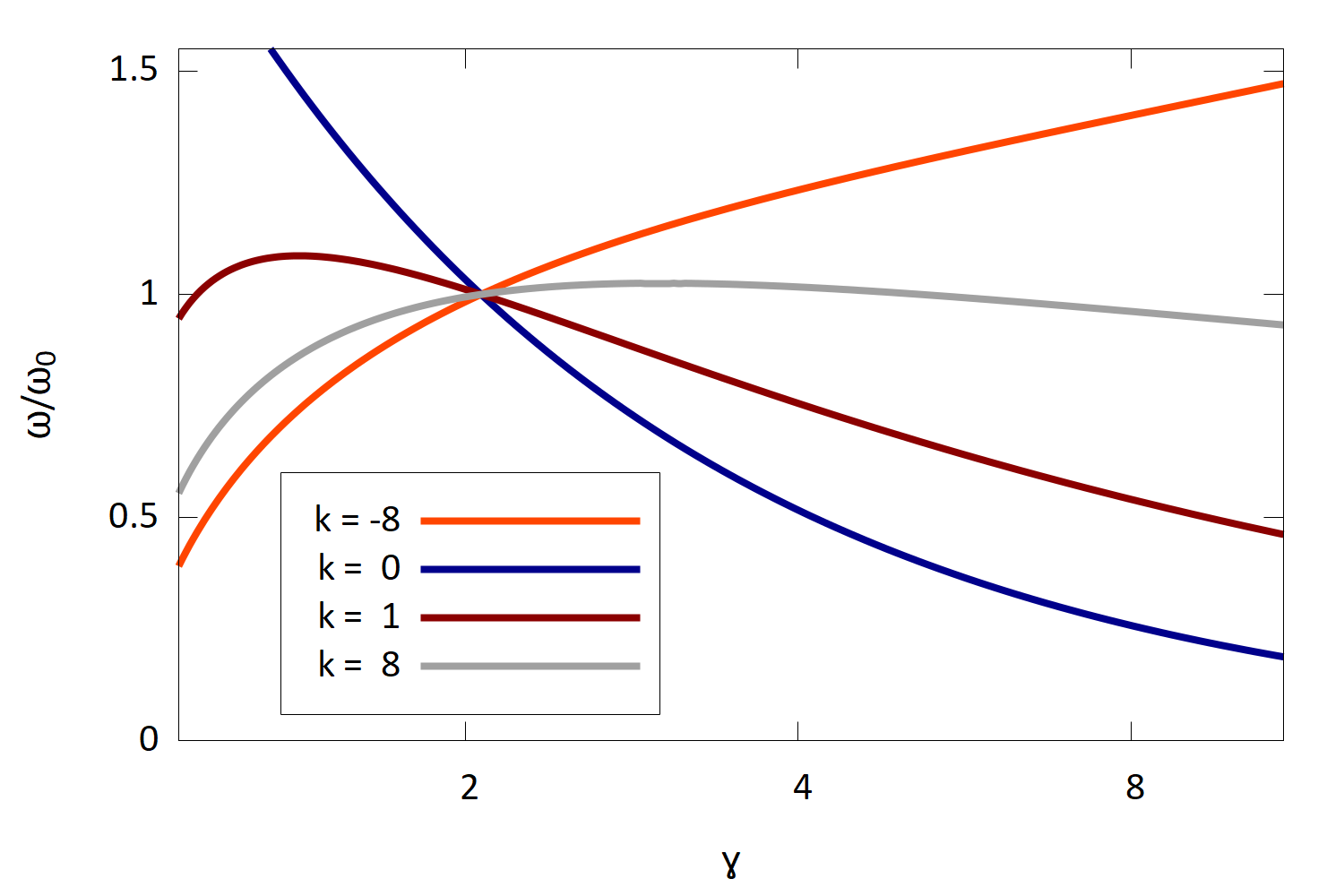}
\caption{Revolution frequency variation as a function of $\gamma$ in a scaling FFA for several field indices $k$. The reference energy is chosen as 1\,GeV.}
\label{fig:ffa_omega}
\end{figure}

In order to obtain small orbit excursions over the range of energies, relatively high values of $k$ are needed. Figure\,\ref{fig:ffa_omega} shows the evolution of Eq. (\ref{omegaffa}) as a function of $\gamma$ for varying $k$ numbers. For positive $k$ the function passes through a maximum at $\gamma_\RM{tr}$ according to Eq. (\ref{gammat}). Using the total energy gain per turn $e U_0$ we can now formulate a relation between synchronous phase $\phi_s$ and the required frequency change per time $\dot{\omega}_\RM{rf}$. The~following relation is equivalent to the expression for the synchro-cyclotron Eq. (\ref{frate2}), but includes now the radial field variation.

\begin{equation}
\frac{e U_0 \sin\phi_s}{E_k + m_0 c^2} = - \frac{2\pi h}{\omega_\RM{rf}^2} \, 
\left( \frac{1 - \gamma^2}{1 - \gamma^2/(k+1)} \right) \dot{\omega}_\RM{rf} \, .
\label{frate3}
\end{equation} 

The need for frequency modulation is still a major disadvantage of these FFA, and schemes to work with fixed frequency are under discussion. One way is to create a large stationary RF bucket that covers the entire range of energies in a cycle. High RF voltages are implemented and the beam is accelerated within a few turns, while stepping along lines of equal potential in the bucket. The other method involves jumps in the harmonic number so that for each turn the fixed RF frequency is an integer multiple of the~beams circulation frequency that varies with increasing beam energy.

Another challenge for scaling FFA is the realisation of the specific radial field shape described by Eq.~(\ref{scaling}). Besides shaping the iron poles of normalconducting magnets, a method for superconducting magnets is the superposition of fields of helical coil layers on a round tube. In this way the required power law can be approximated by a series of multipole fields \cite{witte}.

\subsection{Non-scaling FFA}

Starting with the invention of the cyclotron it was a striking advantage of circular accelerators that the~beam is accelerated over many turns, and the voltage applied for one acceleration step can be kept at a~moderate level. On the downside stability must be ensured for a relatively long acceleration time, and designers had to carefully avoid slow crossing of resonances. As described the radial field shape of scaling FFA allows to keep the betatron tunes constant. However, improvements of the accelerator RF technology made it possible to accelerate beams much more rapid within a few turns, making resonance crossing and variable tunes acceptable. Non-scaling FFA work with varying tunes and can thus be realised using a wider variety of lattice options.

\begin{figure}[h!] 
\centering\includegraphics[width=0.95\textwidth]{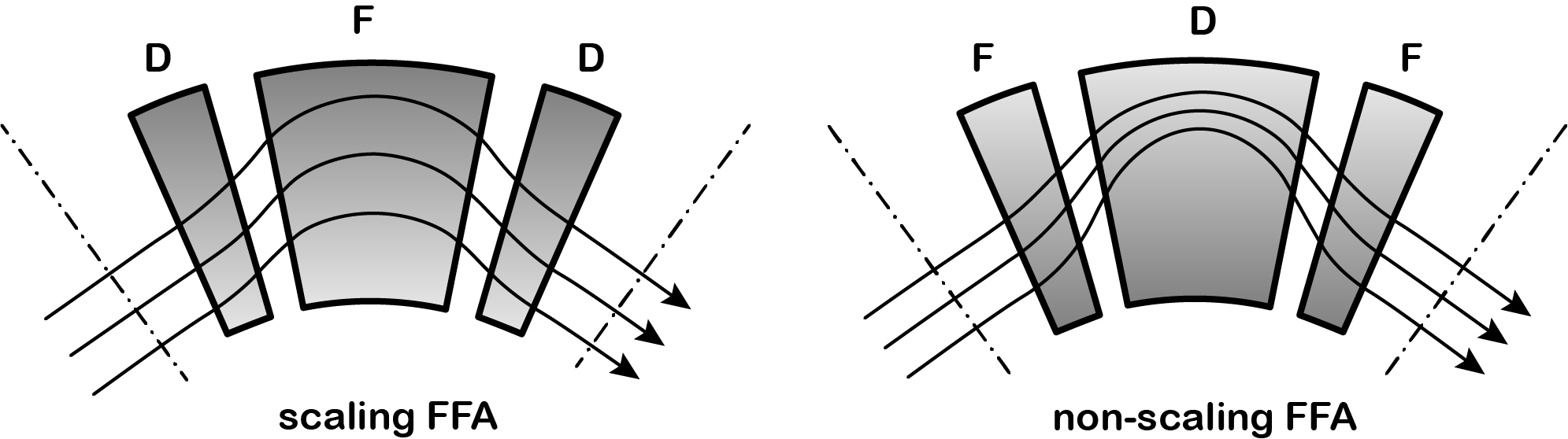}
\caption{Qualitative examples of orbits in a scaling FFA (left) and a non-scaling FFA, after M.\,Craddock.}
\label{fig:ffa_orbits}
\end{figure}

Linear non-scaling FFA use quadrupole fields with linear but strong gradients, avoiding the excitation of non-linear resonances. An example of the cell for such FFA, proposed in Ref. \cite{johnstone}, is shown in the right side of Fig.\,\ref{fig:ffa_orbits}. The cell consists of short focusing quadrupoles and longer defocusing combined function  magnets. In both magnets the absolute field strength decreases towards outer turns. The design example in Ref. \cite{johnstone} allows to accelerate muons from 4\,GeV to 16\,GeV within a physical half aperture of 10--13\,cm, which is a remarkable result for a fixed field accelerator. The partially reverse bending leads to an increase of the average bending radius and the ring size. To obtain the required fields and radial space, comparably large aperture superconducting magnets must be used. In thin lens approximation it can be calculated analytically that the momentum compaction factor depends roughly linear on the beam momentum, and the orbit length is varying quadratically for these types of lattices. In the vicinity of the~minimum the variation is small for relatively large momentum variations. Neighboured RF buckets are slightly shifted in momentum against each other. A fast acceleration process can take place for beams in a channel between the RF buckets, i.e. a region that is normally considered unstable for a storage ring. This method to accelerate a beam with fixed RF frequency is called serpentine channel acceleration. The~method was experimentally demonstrated in EMMA, an electron model FFA in the UK \cite{machida}.

Non-linear non-scaling FFA present a further generalisation of the FFA concept. Even for varying field index, i.e. the non-scaling case, it is still possible to minimize the variation of the tunes over a wide range of momentum by introducing higher order non-linear fields, and this is one of the development goals. Another, perhaps even more important objective is to minimize the circulation time variation for beams of varying momentum, again to avoid cycling of the RF frequency. Using nonlinear fields the~design of practically isochronous FFA rings has been demonstrated for example in Ref. \cite{rees}. An~FFA ring is proposed that accelerates muons from 8 to 20\,GeV in 16 turns. The ring is composed from cells of 5~magnets and has a total length of 1255\,m.

\section{Conclusions}
Both types of accelerators, cyclotron and FFA, belong to the class of fixed field accelerators. Particularly cyclotrons have a long history and represent enabling tools for a variety of research applications, but also medical and industrial applications. For certain purposes where low and medium energy ion beams are needed, the cyclotron presents the most compact and cost efficient solution even 90 years after its invention.

Due to the absence of magnet cycling fixed field accelerators are furthermore a suitable option to provide high average beam intensities as needed for neutron sources or transmutation \cite{rubbia, sheffield}. With 1.4\,MW average beam power the PSI Ring cyclotron is an outstanding example for a high intensity cyclotron. An extrapolation to even higher intensities is proposed in Ref. \cite{stammbach}. Also several FFA studies were developed for high intensity applications. 

The challenges for beam dynamics concern the realisation of transverse stability and resonant acceleration at the same time. While the classical cyclotron is limited at low energies by relativistic effects, the concepts of the synchro-cyclotron and the isochronous cyclotron allow to reach higher energies. In the FFA strong focusing is added and these machines represent the most general concept for fixed field accelerators. We distinguish the scaling FFA with invariant optics properties and the non-scaling FFA, for which a fast acceleration process allows to accept varying tunes. Even with a small number of turns these circular accelerators use costly RF systems in several passes, and thus present often the more economic solution as compared to a LINAC. Another important application of FFA under discussion is the acceleration of muon beams. Due to their limited lifetime muons should be accelerated quickly anyway, and in addition the large acceptance of FFA is a relevant advantage for this purpose. FFA are a topic of intensive research today \cite{icfa} and a broad range of lattice options is studied for different applications.


\begin{thebibliography}{99}

\bibitem{onishenko} L.M.\,Onishchenko, Cyclotrons: A Survey, \emph{Phys. Part. Nuclei} \textbf{39} (2008), p.950, 
\newline \href{https://link.springer.com/article/10.1134/S106377960806004X}{https://link.springer.com/article/10.1134/S106377960806004X}.

\bibitem{seidel_cal} L.\,Calabretta and M.\,Seidel, 50 Years of Cyclotron Development, IEEE Transactions on Nuclear Science, vol.\,63, no.\,2, 2016, pp. 965~--~991,
\href{https://ieeexplore.ieee.org/document/7410111}{https://ieeexplore.ieee.org/document/7410111}.

\bibitem{lawrence} E.O.\,Lawrence and N.E.\,Edlefsen, On the production of high speed protons, \emph{Science} \textbf{72} (1930) 376.

\bibitem{livingston} E.O. Lawrence and M.S. Livingston, The Production of High Speed Light Ions Without the Use of High Voltages, \emph{Phys.\,Rev.} \textbf{40} (1932) 9, \href{http://dx.doi.org/10.1103/PhysRev.40.19}{http://dx.doi.org/10.1103/PhysRev.40.19}.

\bibitem{courant} E.D.\,Courant and H.S.\,Snyder, Theory of the Alternating-Gradient Synchrotron, \emph{Ann. Phys.} \textbf{3} (1958) 1,  \href{http://ab-abp-rlc.web.cern.ch/ab-abp-rlc/AP-literature/Courant-Snyder-1958.pdf}{http://ab-abp-rlc.web.cern.ch/ab-abp-rlc/AP-literature/Courant-Snyder-1958.pdf}.

\bibitem{kleeven} W.\,Kleeven, S.\,Zaremba (IBA), CERN Accelerator School on Accelerators for Medical Application, Cyclotrons: Magnetic Design and Beam Dynamics, 
CERN-2017-004-SP (2017), pp.\,177--239,
\href{https://e-publishing.cern.ch/index.php/CYRSP/article/view/99}{https://e-publishing.cern.ch/index.php/CYRSP/article/view/99}.

\bibitem{veksler} V.\,Veksler, \emph{J. Phys. USSR} \textbf{9(3)} (1945) 153.

\bibitem{mcmillan} E.M.\,McMillan, The Synchrotron—A Proposed High Energy Particle Accelerator, \emph{Phys. Rev.} \textbf{68(5-6)} (1945) 143L,
\href{http://dx.doi.org/10.1103/PhysRev.68.143}{http://dx.doi.org/10.1103/PhysRev.68.143}. 

\bibitem{kleeven2} W.\,Kleeven et al, The IBA Superconducting Synchrocyclotron Project S2C2, Proc. 20th Int. Conf. on Cyclotrons and Their Applications, Vancouver, 2013, pp.\,115--119, \newline\href{http://accelconf.web.cern.ch/CYCLOTRONS2013/papers/mo4pb02.pdf}{http://accelconf.web.cern.ch/CYCLOTRONS2013/papers/mo4pb02.pdf}.

\bibitem{thomas} L.H.\,Thomas, The Paths of Ions in the Cyclotron, \emph{Phys. Rev.} \textbf{54} (1938) pp.\,580--598, \newline
\href{http://prola.aps.org/pdf/PR/v54/i8/p580_1}{http://prola.aps.org/pdf/PR/v54/i8/p580\_1}.

\bibitem{bi} Y.J.\,Bi \emph{et al.}, Towards quantitative simulations of high power proton cyclotrons, \emph{Phys. Rev. Spec. Top. Accel. Beams} \textbf{14} (2011) 054402, 
\href{http://link.aps.org/doi/10.1103/PhysRevSTAB.14.054402}{http://link.aps.org/doi/10.1103/PhysRevSTAB.14.054402}.

\bibitem{seidel3} M.\,Seidel, Injection and Extraction in Cyclotrons, CERN Accelerator School on Beam Injection, Extraction and Transfer, CERN-2018-008-SP (2017) pp.\,151--162, \newline
\href{https://doi.org/10.23730/CYRSP-2018-005.151}{https://doi.org/10.23730/CYRSP-2018-005.151}.

\bibitem{calabretta} L.\,Calabretta \emph{et al.}, A multi-megawatt cyclotron complex to search for CP violation in the neutrino sector, Proc. Cyclotrons and Their Applications 2010, Lanzhou, China, 2010, p.\,299, \newline
\href{http://accelconf.web.cern.ch/AccelConf/Cyclotrons2010/papers/tua1cio01.pdf}{http://accelconf.web.cern.ch/AccelConf/Cyclotrons2010/papers/tua1cio01.pdf}.

\bibitem{joho} W.\,Joho, High intensity problems in cyclotrons, Proc. 5th Int.
Conf. on Cyclotrons and Their Applications, Caen, 1981, pp.\,337--347,
\href{http://accelconf.web.cern.ch/c81/papers/ei-03.pdf}{http://accelconf.web.cern.ch/c81/papers/ei-03.pdf}.

\bibitem{handbook} A.\,Chao and M.\,Tigner (Eds.), Handbook of Accelerator Physics and Engineering (World Scientific, Singapore, 1999), Chapter 1.6.4.

\bibitem{jianjun} J.J.\,Yang \emph{et al.}, Beam dynamics in high intensity cyclotrons including neighboring bunch effects: Model, implementation, and application, \emph{Phys. Rev. Spec. Top. Accel. Beams} \textbf{13} (2010) 064201, \newline
\href{http://link.aps.org/doi/10.1103/PhysRevSTAB.13.064201}{http://link.aps.org/doi/10.1103/PhysRevSTAB.13.064201}.

\bibitem{dutto} G.\,Dutto \emph{et al.}, TRIUMF high intensity cyclotron development for ISAC, pp.\,82--86,
Proc. 17th Int. Conf. on Cyclotrons and Their Applications, Tokyo, 2004, \newline
\href{http://accelconf.web.cern.ch/AccelConf/c04/data/CYC2004_papers/18C3.pdf}{http://accelconf.web.cern.ch/AccelConf/c04/data/CYC2004\_papers/18C3.pdf}.

\bibitem{kase} M.\,Kase \emph{et al.}, Present status of the RIKEN Ring cyclotron, Proc. 17th Int. Conf. on Cyclotrons and Their Applications, Tokyo, 2004, pp.\,160--162, \newline
\href{http://accelconf.web.cern.ch/AccelConf/c04/data/CYC2004_papers/18P24.pdf}{http://accelconf.web.cern.ch/AccelConf/c04/data/CYC2004\_papers/18P24.pdf}.

\bibitem{okuno} H.\,Okuno \emph{et al.}, Magnets for the RIKEN superconducting RING cyclotron, 
Proc. 17th Int. Conf. on Cyclotrons and Their Applications, Tokyo, 2004, pp.\,373--377, \newline
\href{http://accelconf.web.cern.ch/AccelConf/c04/data/CYC2004_papers/20B1.pdf}{http://accelconf.web.cern.ch/AccelConf/c04/data/CYC2004\_papers/20B1.pdf}.

\bibitem{willax} H.\,Willax, Proposal for a 500~MeV isochronous cyclotron with ring magnet, 
Proc. Int. Conf. on Sector-Focused Cyclotrons, Geneva, 1963, p.\,386, 
\href{http://epaper.kek.jp/c63/papers/cyc63h04.pdf}{http://epaper.kek.jp/c63/papers/cyc63h04.pdf}.

\bibitem{seidel} M.\,Seidel \emph{et al.}, Production of a 1.3~MW proton beam at PSI, Proc. IPAC'10, Kyoto, 2010, pp.\,1309-1313,
\href{http://accelconf.web.cern.ch/Accelconf/IPAC10/papers/tuyra03.pdf}{http://accelconf.web.cern.ch/Accelconf/IPAC10/papers/tuyra03.pdf}.

\bibitem{craddock} M.A.\,Craddock, presentation at the FFAG'08 workshop, Manchester (2008), \newline\href{https://www.cockcroft.ac.uk/events/FFAG08/presentations/Craddock/Thomas-FFAG.pdf}{https://www.cockcroft.ac.uk/events/FFAG08/presentations/Craddock/Thomas-FFAG.pdf}.

\bibitem{symon} K.R.\,Symon, D.W.\,Kerst, L.W.\,Jones, L.J.\,Laslett and K.M.\,Terwilliger, \emph{Phys. Rev.} \textbf{103} (1956) 1837.

\bibitem{gantry} D.\,Trbojevic, B.\,Parker, E.\,Keil, A.M.\,Sessler, Carbon/proton therapy: A novel gantry design, Phys. Rev. ST Accel. Beams (2007) p.\,053503,\newline
\href{https://link.aps.org/doi/10.1103/PhysRevSTAB.10.053503}{https://link.aps.org/doi/10.1103/PhysRevSTAB.10.053503}. 

\bibitem{witte} H.\,Witte et al, PAMELA Magnets - Design and Performance, p.\,301, \newline
\href{https://accelconf.web.cern.ch/PAC2009/papers/mo6pfp073.pdf}{https://accelconf.web.cern.ch/PAC2009/papers/mo6pfp073.pdf}.

\bibitem{johnstone} C.\,Johnstone, W.\,Wan, A.\,Garren, Fixed field circular accelerator designs, 
Proc. PAC'99, New York, 1999, p.\,3068, \href{https://accelconf.web.cern.ch/p99/PAPERS/THP50.PDF}{https://accelconf.web.cern.ch/p99/PAPERS/THP50.PDF}.

\bibitem{machida} S.\,Machida, Acceleration in the linear non-scaling fixed-field alternating-gradient accelerator EMMA, Nat.\,Phys., vol. 8, no. 3, 2012, pp.\,243--247.

\bibitem{rees} G.H.\,Rees, FFAG’04 (2004); FFAG’05 (2005); and \cite{icfa}.

\bibitem{rubbia} C. Rubbia \emph{et al.}, An energy amplifier for cleaner and inexhaustible nuclear energy production driven by a particle beam accelerator, CERN/AT/93-47 (ET) (1993), \newline
\href{http://cdsweb.cern.ch/record/256520/files/at-93-047.pdf}{http://cdsweb.cern.ch/record/256520/files/at-93-047.pdf}.

\bibitem{sheffield} R.\,Sheffield, Utilization of accelerators for transmutation and energy production, Proc. HB2010, Morschach, Switzerland, 2010, pp.\,1--5, \newline
\href{http://accelconf.web.cern.ch/AccelConf/HB2010/papers/moia01.pdf}{http://accelconf.web.cern.ch/AccelConf/HB2010/papers/moia01.pdf}.

\bibitem{stammbach} T.\,Stammbach \emph{et al.}, The feasibility of high power cyclotrons, \emph{Nucl. Instrum. Methods Phys. Res. B} \textbf{113} (1996) pp.\,1--7,
\href{http://www.sciencedirect.com/science/article/pii/0168583X95013776}{http://www.sciencedirect.com/science/article/pii/0168583X95013776}.

\bibitem{icfa} ICFA-Beam Dynamics Newsletter, \textbf{43}, 74 (2007), \href{http://www.icfa-bd.org/Newsletter43.pdf}{http://www.icfa-bd.org/Newsletter43.pdf}.

\end{thebibliography}
\end{document}